\documentclass[11pt]{article}
\usepackage{graphics,cite,amssymb,epsfig,float}
\usepackage[usenames,dvips]{color}
\usepackage{amsmath, mathbbol}

\usepackage{multirow}
\def\lsim{\;\raise0.3ex\hbox{$<$\kern-0.75em\raise-1.1ex\hbox{$\sim$}}\;}
\def\gsim{\;\raise0.3ex\hbox{$>$\kern-0.75em\raise-1.1ex\hbox{$\sim$}}\;}

\def\ben{\begin{enumerate}}  \def\een{\end{enumerate}}
\def\bit{\begin{itemize}}    \def\eit{\end{itemize}}
\def\beq{\begin{equation}}   \def\eeq{\end{equation}}
\def\ba{\begin{array}}       \def\ea{\end{array}}
\def\bea{\begin{eqnarray}}   \def\eea{\end{eqnarray}}

\begin{document}

\setcounter{footnote}{0}
\vspace*{-2.5cm}
\begin{flushright}
LPT Orsay 13-54 \\
PCCF RI 13-05\\
IFT-UAM/CSIC-13-118 \\
FTUAM-13-31

\vspace*{2mm}
\end{flushright}
\begin{center}
\vspace*{5mm}

\vspace{1cm}
{\Large\bf 
Sterile neutrinos in leptonic and semileptonic decays}\\
\vspace{1cm}

{\bf A. Abada$^{a}$, A.M. Teixeira$^{b}$, A. Vicente$^{a,c}$ and
  C. Weiland$^{d,e}$ }

\vspace*{.5cm} 
$^{a}$ Laboratoire de Physique Th\'eorique, CNRS -- UMR 8627, \\
Universit\'e de Paris-Sud 11, F-91405 Orsay Cedex, France
\vspace*{.2cm} 

$^{b}$ Laboratoire de Physique Corpusculaire, CNRS/IN2P3 -- UMR 6533,\\ 
Campus des C\'ezeaux, 24 Av. des Landais, F-63171 Aubi\`ere Cedex, France
\vspace*{.2cm}

$^{c}$ IFPA, Dep. AGO, Universit\'e de Li\`ege, \\
Bat B5, Sart-Tilman B-4000 Li\`ege 1, Belgium
\vspace*{.2cm}

$^{d}$ Departamento de F\'isica Te\'orica, Universidad Aut\'onoma de Madrid,\\
Cantoblanco, Madrid 28049, Spain

\vspace*{.2cm}
$^{e}$ Instituto de F\'isica Te\'orica UAM/CSIC,\\
Calle Nicol\'as Cabrera 13-15, Cantoblanco, Madrid 28049, Spain

\end{center}

\vspace*{10mm}
\begin{abstract}

We address the impact of a modified $W \ell \nu$ coupling on a wide
range of observables, such as $\tau$ leptonic and mesonic decays,
leptonic decays of pseudoscalar mesons, as well as semileptonic meson
decays. In particular, we concentrate on deviations from
lepton flavour universality, focusing on the ratios $R_{P} = \Gamma (P
\to \ell \nu) / \Gamma (P \to \ell' \nu)$, with $P=K, \pi, D, D_s$,
$R(D)={\Gamma (B^+ \to D \tau^+
  \nu)}/{\Gamma (B^+ \to D\ell^+ \nu)}$, $R_\tau={\Gamma (\tau\to
  \mu\nu\nu)}/{\Gamma (\tau\to e\nu\nu)}$, $R^{\ell
  \tau}_P=\Gamma(\tau\to P\nu)/\Gamma(P\to \ell\nu)$, and 
$\text{BR}(B \to \tau \nu)$.  We further
consider leptonic gauge boson decays, such as $W\to \ell \nu $
and $Z \to \nu \nu$.  For all the above observables, we
provide the corresponding complete analytical expressions, derived for
the case of massive neutrinos.  Working in the framework of the
Standard Model extended by additional sterile fermions, which mix with
the active (left-handed) neutrinos, we numerically 
study the impact of active-sterile mixings on the above mentioned observables.
\end{abstract}

\vspace*{3mm}

\section{{Introduction}}

In order to account for neutrino masses and
mixings, the Standard Model (SM) 
can be extended with new  sterile fermionic
states, such as right-handed neutrinos.
Sterile states are present in several neutrino mass models, and their
existence is also strongly motivated by current data from reactor
experiments, cosmology, as well as indications from large scale structure
formation~\cite{Kusenko:2009up,Abazajian:2012ys}.
In these frameworks, leptonic charged currents can be modified due to 
the mixings of the sterile neutrinos with the active
left-handed ones. The SM  
flavour-conserving term in the lepton weak charged current Lagrangian
is modified as 
\begin{equation}\label{eq:cc-lag}
- \mathcal{L}_\text{cc} = \frac{g}{\sqrt{2}} U^{ji} 
\bar{\ell}_j \gamma^\mu P_L \nu_i  W_\mu^- + \, \text{c.c.}\,,
\end{equation}
where $U$ is a generic leptonic mixing matrix,
$i = 1, \dots, n_\nu$ denotes the physical neutrino states
and $j = 1, \dots, 3$ the flavour of the charged leptons. 
In the case of three neutrino generations,  $U$ corresponds
to the unitary PMNS matrix, $U_\text{PMNS}$. 
The mixing between the left-handed leptons, here denoted by $\tilde
U_\text{PMNS}$, 
now corresponds to a $3 \times 3$ block of $U$, which can be parametrised as
\begin{equation}\label{eq:U:eta:PMNS2}
U_\text{PMNS} \, \to \, \tilde U_\text{PMNS} \, = \,(\mathbb{1} - \eta)\, 
U_\text{PMNS}\,,
\end{equation}
where the  matrix $\eta$ contains the deviation of $\tilde
U_\text{PMNS}$ from unitarity~\cite{Schechter:1980gr,Gronau:1984ct}.

The active-sterile mixings and the departure from unitarity of 
$\tilde U_\text{PMNS}$
can have an impact on several observables, inducing deviations from
SM predictions, such as violation  of lepton flavour 
universality (LFU)~\cite{Shrock, Nardi:1994iv,Abada:2012mc}, 
enhanced lepton flavour violating (LFV) 
processes~\cite{Ilakovac:1994kj,Deppisch:2004fa} and new contributions
to different 
low-energy rare decays.
 
In this work we address the impact of the modified charged current
vertex on several observables, whose
dominant SM contribution arises from tree-level $W$ exchange. This is the
case of decays with one or two neutrinos in the final state, as for
example $\tau$ leptonic and mesonic decays, 
leptonic $ \pi,\ K,\ D, \ D_s,\ B$ decays and semileptonic meson
decays, like $B\to D\ell \nu$.  
We also consider leptonic gauge boson
decays, such as $W\to \ell \nu $ and $Z \to \nu \nu$. 
Despite the fact that the hadronic sector can also be affected by some 
underlying New Physics (NP) contributions, in our analysis we
decorrelate these effects, assuming that all NP effects are encoded
in the modified leptonic weak
current vertices, due to the presence of extra sterile neutrinos.
To do so, and for all the observables mentioned above, we derive
the corresponding complete analytical expressions in this context, in
particular, fully accounting for massive charged and neutral fermions
as well as mixing in the lepton sector. 

The deviations from unitarity as well as the possibility of having
the sterile states as final decay products might induce 
departures from the SM theoretical expectations. 
Due to these  potential contributions, these
frameworks are severely constrained:  any realisation must comply with
a number of laboratory bounds, electroweak (EW) precision tests and
cosmological constraints, among others.

The modified $W \ell \nu$ vertex, and the associated 
new contributions to the different observables mentioned above
can be found in several scenarios with additional singlet
states, as is the case of the $\nu$SM~\cite{Asaka:2005an}, the low-scale
type-I seesaw~\cite{Ibarra:2010xw} and the Inverse Seesaw
(ISS)~\cite{Mohapatra:1986bd}, among other possibilities. 
For the purpose of our numerical analysis, it is convenient to consider a
specific seesaw realisation which consists of an extension of the SM
field content  by sterile neutrinos. As done 
in a first study devoted to LFU violation in kaon and pion
leptonic decays~\cite{Abada:2012mc}, we consider here the ISS, which has the
appealing feature of naturally having large Yukawa couplings and
a comparatively light sterile spectrum at the same time, thus increasing the
active-sterile mixing. It is nevertheless worth pointing out that 
our results are quite general, since they only depend on the modified 
$W \ell \nu$ vertex. 
Therefore, and although we choose a specific framework, 
the qualitative conclusions here derived should in principle hold for other models.

Our work is organised as follows: in Section~\ref{sec:SMnusterile}
we address in  detail the departure from unitarity of the
$\tilde U_\text{PMNS}$ matrix, focusing on the different constraints arising
from neutrino data, electroweak observables, laboratory measurements
and cosmological observations. In Section~\ref{sec:observables}, we
discuss the observables considered, presenting analytical
formulae, and we summarise the corresponding SM expectations and
experimental status. The numerical results (for the specific
realisation of the ISS considered) are collected and discussed in Section~\ref{sec:results},
while our concluding remarks are given in Section~\ref{sec:concs}.

\section{The SM extended by fermionic gauge singlets }\label{sec:SMnusterile}

One of the simplest extensions of the SM allowing to accommodate
massive neutrinos consists in the introduction of right-handed states
$\nu_R$, singlets under the SM gauge group. The SM mass Lagrangian is
enlarged with a Dirac mass term $m_D \bar \nu_R \nu_L$, and should
lepton number violation be allowed, with a Majorana mass term $m_M \bar
\nu_R^c \nu_R$.  Within this class of models, the standard type-I
seesaw~\cite{Minkowski:1977sc,seesaw,Mohapatra:1979ia} is an appealing
framework, where a natural explanation for the smallness of neutrino
masses can be found by assuming that the Majorana masses of the
right-handed neutrinos are large, leading to a suppression of $m_\nu
\sim m_{D}^2/m_M$.
 
However, the large mass scales usually involved, typically much larger
than the electroweak scale, imply that no direct experimental tests of
the standard type-I seesaw model are possible.  Low-scale seesaw
models, in which the new singlet fermions are lighter, with masses
around the electroweak scale, are more attractive from a
phenomenological point of view.  In this case, the new states can be
produced in collider or low-energy experiments and their
contributions to physical processes can be sizable. In our work we
will consider this type of models\footnote{An interesting realisation
  of low-scale seesaw models is the so-called Inverse
  Seesaw~\cite{Mohapatra:1986bd}, which will be briefly reviewed in
  Section~\ref{Subsec:ISS} and subsequently used in the numerical
  analysis.}.

\subsection{Impact for charged currents: the ${\pmb{W \ell \nu}}$ vertex}
In the framework of the SM extended to accommodate massive neutrinos, the lepton weak
charged current Lagrangian is given by 
$-{g}/{\sqrt{2}}\, J^\mu\, W_\mu^- + \,\text{c.c.}\,,$ where 
$J^\mu = 
\bar{\ell}\, U\, \gamma^\mu\, P_L\, \nu$, with $P_L =
(\mathbb{1} - \gamma_5)/2$ and
\begin{equation}
U\, =\, V^\dagger\, U_\nu\,.
\end{equation}
In the above, $V$ and $U_\nu$ are unitary transformations that relate the physical
$\ell$ and $\nu$ states to the gauge eigenstates $\ell'$ and $\nu'$ as
\begin{equation}
\ell^\prime\, =\, V\, \ell \,,\qquad \qquad \nu^\prime \,=\, U_\nu\, \nu\,,
\end{equation}
and the matrix $U$ is thus the leptonic mixing matrix (the 
analog of the CKM matrix in the quark sector); just 
as in the quark sector, where the flavour structure of the
CKM matrix leads to a very rich phenomenology, the leptonic mixing
matrix also has an impact on many observables related to lepton
flavour.

The above discussion is generic, and holds in scenarios with
additional singlet neutrinos. 
However, since only left-handed leptons participate in the charged
interaction, in this case $U$ 
is a rectangular matrix which can be written as
\begin{equation}
U \,= \,\left(\tilde U_\text{PMNS} \, , \, U_{AS} \right) \, ,
\end{equation}
where $\tilde U_\text{PMNS}$ is a $3 \times 3$ matrix and $U_{AS}$ is
a $3 \times (n_\nu-3)$ matrix, with $n_\nu$ the total number of
neutrino states. While the rows of $U$ are indeed 
unit vectors ($U\, U^\dagger = \mathbb{1}$), 
the $\tilde U_\text{PMNS}$ and $U_{AS}$
submatrices are not unitary~\cite{Schechter:1980gr}.

One can easily interpret the matrices $\tilde U_\text{PMNS}$ and
$U_{AS}$. In the case of three neutrino generations 
(no additional sterile states), $U$
corresponds to the unitary PMNS matrix and thus one can identify $\tilde
U_\text{PMNS} = U_\text{PMNS}$. However, in general, the mixing among
the left-handed leptons is given by a non-unitary $\tilde
U_\text{PMNS}$, usually parametrised as already introduced 
in Eq.~(\ref{eq:U:eta:PMNS2}), 
\begin{equation}
\tilde U_\text{PMNS} \, = \,(\mathbb{1} - \eta)\, U_\text{PMNS}\,.
\nonumber
\end{equation}
Finally, the matrix $U_{AS}$ contains information about the mixing between
the active neutrinos and the sterile singlet states.

\subsection{Constraints from neutrino data}
\label{subsec:nudata}
Any neutrino motivated extension of the SM must accommodate
oscillation
data~\cite{Tortola:2012te,Fogli:2012ua,GonzalezGarcia:2012sz}. In our
analysis, we consider both possible (normal and inverted) hierarchies
for the light neutrino spectrum and take the current best-fit results
on the oscillation parameters as obtained
in~\cite{Tortola:2012te}. For a normal hierarchy this implies the
following values
\begin{align}\label{eq:neutrino.data.NH}
\sin^2 \theta_{12}\,=\, 0.32\,, 
\quad
\sin^2 \theta_{23}\,=\, 0.427\,, 
\quad
\sin^2 \theta_{13}\,=\, 0.0246\,, \nonumber \\
\Delta m^2_{21}\,=\, 7.62\times 10^{-5} \mathrm{eV}^2\,, 
\quad
|\Delta m^2_{31}|\,=\, 2.55\times 10^{-3} \mathrm{eV}^2\, . 
\end{align}
Note that we have taken the local minimum for $\theta_{23}$ in the
first octant, in agreement with~\cite{Fogli:2012ua}. On the other hand,
for an inverted hierarchy we have
\begin{align}\label{eq:neutrino.data.IH}
\sin^2 \theta_{12}\,=\, 0.32\,, 
\quad
\sin^2 \theta_{23}\,=\, 0.6\,, 
\quad
\sin^2 \theta_{13}\,=\, 0.025\,, \nonumber \\
\Delta m^2_{21}\,=\, 7.62\times 10^{-5} \mathrm{eV}^2\,, 
\quad
|\Delta m^2_{31}|\,=\, 2.43\times 10^{-3} \mathrm{eV}^2\, . 
\end{align}
Although no data on the CP violating phases is available, we have also investigated the effect of the Dirac phase in our analysis. 

As recently pointed out~\cite{Qian:2013ora},
a combination of solar neutrino experiments, medium-baseline and
short-baseline reactor antineutrino experiments could allow to
perform the first direct unitarity test of the PMNS matrix. 

Bounds on the non-unitarity matrix $\eta$, defined in
Eq.~(\ref{eq:U:eta:PMNS2}), were derived using Non-Standard
Interactions~\cite{Antusch:2008tz}. However, they were obtained by
means of an effective theory approach, and thus their application in
our numerical study will be limited to the cases in which the latter
approach is valid.

\subsection{Constraints from EW precision tests}
Let us begin by briefly commenting on the constraints
derived from global fits to electroweak precision data. In the
presence of singlet neutrinos, electroweak precision constraints were
first addressed in~\cite{delAguila:2008pw} and recently studied
in~\cite{Akhmedov:2013hec,Basso:2013jka}. In~\cite{delAguila:2008pw}, an effective
approach was used, and thus these constraints will only be considered
in cases with multi-TeV singlet states.  Our numerical results, which
will be presented in Section~\ref{sec:results}, are in agreement with
the results of~\cite{Akhmedov:2013hec,Basso:2013jka}.

\subsubsection*{$\pmb{Z}$ invisible decay width}
The comparison of the SM prediction of the $Z$ invisible decay width
to the LEP measurement~\cite{Beringer:1900zz}, 
\begin{align}
&\Gamma_\text{SM}(Z \to \nu \nu) \,=\, (501.69 \pm 0.06) \, \text{MeV} \,,
\label{eq:Znunu:SM}\\
&\Gamma_\text{Exp}(Z \to \nu \nu) \,=\, (499.0 \pm 1.5) \, \text{MeV} \,, 
\label{eq:Znunu:Exp}
\end{align}
suggests that the experimental value is $\sim 2 \,\sigma$ below the
theoretical expectation of  the SM. In order to investigate if the
presence of sterile fermions could have an impact on the decay width
$\Gamma(Z \to \nu \nu)$, one needs to consider the latter decay in the
general case of massive Majorana neutrinos.  The invisible $Z$ decay width
reads
\begin{eqnarray}
\Gamma(Z \to \nu \nu) &=& \sum_{i,j} \Delta_{ij}
\Gamma_{VFF}(m_Z,m_{\nu_i},m_{\nu_j},b_L^{ij},b_R^{ij})\ ,
\label{eq:Znunu:sum}
\end{eqnarray}
where $i ,j=1, \dots, N_\text{max}$ ($N_\text{max}$ corresponding to the heaviest
$\nu$ which is kinematically allowed). The function $\Gamma_{VFF} \equiv
\Gamma_{VFF}(m_V,m_{F_1},m_{F_2},b_L,b_R)$ is given by
\begin{align}
\!\!\!\Gamma_{VFF}\, &=\,
\frac{{\lambda^{1/2}(m_V,m_{F_1} ,m_{F_2})}}{48\, \pi\, m_V^3} \nonumber \\
& \times \left[\left(|b_L|^2+|b_R|^2\right) 
\left(-\frac{\left(m_{F_1}^2-m_{F_2}^2\right)^2}{m_V^2}-m_{F_1}^2-m_{F_2}^2+2
m_V^2\right)+12 m_{F_1} 
   m_{F_2} \text{Re}\left(b_L b_R^*\right)\right] . \label{eq:GammaVFF}
\end{align}
In the above, the kinematical function $\lambda(a,b,c)$ is defined as 
\bea\label{lambda.function}
\lambda(a,b,c) \, = \, (a^2 - b^2 -c^2)^2 -
4\,b^2\,c^2\,,
\eea
and the couplings $b_{L,R}$ are 
\begin{eqnarray}\label{eq:Znunu:couplings}
b_L^{ij} &=& 2^{1/4} m_Z \sqrt{G_F} \sum_{a=1}^3 U_{ai}^*  U_{aj}\ , 
\nonumber \\
b_R^{ij} &=& - \left(b_L^{ij}\right)^*\,.
\end{eqnarray}
Finally, $\Delta_{ij} = 1 - \frac{1}{2} \delta_{ij}$ is a factor which 
accounts for the Majorana nature of the neutrinos.

Since the latter expressions depend on the entries of the mixing matrix $U$, 
and given that the sum in Eq.~(\ref{eq:Znunu:sum}) involves all neutrino
states that are  kinematically allowed, the $Z$ invisible decay width is an important constraint 
 that any
model involving fermionic gauge singlets should satisfy.

\subsection{Other constraints}

Sterile neutrinos can be produced in meson decays such as $\pi^\pm \to
\mu^\pm \nu$, with rates dependent on their mixing with the active
neutrinos. Negative searches for monochromatic lines in the muon
spectrum can be translated into bounds for $m_{N_s} - \theta_{\mu s}$
combinations, where $m_{N_s} $ is the mass of one of the sterile
states and $\theta_{\mu s}$ parametrises the active-sterile
mixing~\cite{Atre:2009rg,Kusenko:2009up}. All experimental searches
performed so far have led to negative results, allowing  to set
stringent limits for sterile neutrinos with masses in the MeV-GeV
range and fuelling plans for future experiments~\cite{Lello:2012gi}.

\medskip
Unless the active-sterile mixings are negligible, the modified $W \ell
\nu$ vertex may also contribute to LFV processes, with rates
potentially larger than current bounds. The radiative muon decay $\mu
\to e \gamma$, searched for by the MEG experiment~\cite{Adam:2013mnn},
provides the most stringent constraint. The rate induced by sterile
neutrinos\footnote{We
  have assumed a dipole dominated LFV phenomenology. In this case $\mu
  \to e \gamma$ is the most constraining LFV observable. However, it
  has been recently pointed out that in low-scale seesaw models, the
  dominant contributions might come from (non-supersymmetric) box
  diagrams~\cite{Ilakovac:2009jf,Alonso:2012ji,Dinh:2012bp,Ilakovac:2012sh}. In
  this case, the expected future sensitivity of $\mu- e$ conversion
  experiments can also play a relevant r\^ole in detecting and/or
  constraining sterile neutrino scenarios. Similarly, supersymmetric
  models may have dominant contributions beyond the
  dipole one~\cite{Abada:2012cq}.} should
obey~\cite{Ilakovac:1994kj,Deppisch:2004fa}
\begin{equation}\label{eq:BR:muegamma:sterile}
\text{BR}(\mu \to e \gamma) \,=\,  
\frac{\alpha_W^3 \,s_W^2 \,m_\mu^5}{256 \,\pi^2\, m_W^4\, \Gamma_\mu} |H_{\mu e}|^2
\leq 5.7 \times 10^{-13}\, ,
\end{equation}
where $H_{\mu e} = \sum_i U^{2i} U^{1i \, *} G_\gamma \left(
\frac{m_{\nu,i+3}^2}{m_W^2} \right)$,  $G_\gamma$ being the associated
loop function and $U$  the mixing matrix defined in
Eq. \eqref{eq:cc-lag}. In addition, $\alpha_W$ and $s_W = \sin
\theta_W$ denote the weak coupling and mixing angle, respectively,
while $\Gamma_\mu$ corresponds to the total muon width.

\medskip
At higher energies, constraints on sterile neutrinos can also be
derived from Higgs decays.  LHC data already provides
some important bounds when the sterile states are slightly below $125$
GeV, due to the potential Higgs decays to left- and right-handed
neutrinos. This has been recently studied in
\cite{BhupalDev:2012zg,Cely:2012bz,Bandyopadhyay:2012px}.

Finally, in the presence of lepton number violating (LNV)
interactions, as is the case for singlet neutrinos with Majorana
masses, new processes are possible. Although neutrinoless double beta
decay \cite{Deppisch:2012nb} remains the key observable (the
  most recent results on neutrinoless double beta decay having been 
  obtained by the GERDA experiment~\cite{Agostini:2013mzu}), the LHC
is beginning to be competitive, as demonstrated by the
phenomenological studies of~\cite{Helo:2013dla,Helo:2013ika} and
recent CMS results~\cite{Chatrchyan:2012fla}. However, we have not
taken these LNV processes into account, since they are not correlated
with the observables of interest for our study\footnote{As mentioned
  in Section~\ref{subsec:nudata}, we take vanishing Majorana
  phases. The latter have no impact on the observables studied in this
  work but may possibly be used as
  degrees of freedom to lower the rates for LNV processes below
  current bounds.}.

\subsection{Constraints from cosmology}
Under the assumption of a standard cosmology, the most constraining
bounds on sterile neutrinos with a mass below the TeV arise from a
wide variety of cosmological
observations~\cite{Smirnov:2006bu,Kusenko:2009up}. Sterile neutrinos
can constitute a non-negligible fraction of the dark matter of the
Universe and thus influence structure formation, which is constrained
by Large Scale Structure and Lyman-$\alpha$ data.
Active-sterile mixing also induces the radiative decays $\nu_i \to \nu_j
\gamma$, well constrained by cosmic X-ray searches. Lyman-$\alpha$
limits, the existence of additional degrees of freedom at the epoch of
Big Bang Nucleosynthesis, and Cosmic Microwave Background data (among others), 
allow to set additional bounds in the $m_{N_s} - \theta_{is}$
plane. However, all the above cosmological bounds can be evaded if a
non-standard cosmology is considered, for example in scenarios with a
low reheating temperature~\cite{Gelmini:2008fq}, or when sterile
neutrinos couple to  a dark sector~\cite{Dasgupta:2013zpn}.  In our numerical
analysis we will allow for the violation of the latter bounds in some scenarios,
explicitly stating it.

\section{Observables}\label{sec:observables}

We now proceed to derive the new contributions to a number of
observables involving the  $W \to \ell \nu$ vertex. 
Some of these expressions have also been derived in~\cite{Gorbunov:2007ak}.
Although for many of the considered observables there is a good
agreement between the SM expectations and experimental measurements,
for some others there is a manifest tension between theoretical predictions
and experimental results.  We will explore how extending the SM by sterile
neutrinos might contribute to alleviate some of the latter tensions. 
This will depend on the sterile neutrino  masses and  on their
mixings with the active neutrinos. Should  the sterile fermions be light, they
can be kinematically available as final states of a given decay,
leading in some cases to a further enhancement from phase space effects.
In this work we thus address (when possible)
observables which allow to reduce the need of hadronic input.  

\subsection{$\pmb{W \to \ell \nu}$ decays}\label{sec:Wdecays}
We first consider the observable most directly affected by the sterile 
fermions -  $W$ leptonic decays, BR($W \to \ell_i \nu$).
The width of the $W \to \ell_i \nu$ decay is given by 
\begin{equation}\label{eq:Wellnu:sum}
\Gamma(W \to \ell_i \nu) = \sum_{j=1}^{N_\text{max}^{(\ell_i)}} 
\Gamma_{VFF}(m_W,m_{\ell_i},m_{\nu_j},a_L^{ij},0) \ , 
\end{equation}
where the functions $\Gamma_{VFF}$ and $\lambda(m_W,m_{F_1},m_{F_2})$
are given in Eqs. \eqref{eq:GammaVFF} and \eqref{lambda.function},
respectively.  The couplings $a_L$ are defined as
\begin{equation}\label{eq:Wellnu:couplings}
a_L^{ij} \,=\, 2^{3/4}\, m_W\, \sqrt{G_F}\, U_{ij} \,.
\end{equation}
Not necessarily all $\nu_j$ can be final products of the decay.  We denote by
$N_\text{max}^{(\ell_i)}$ the $N^\text{th}$ heaviest neutrino mass
eigenstate which is kinematically allowed when the lepton produced is
$\ell_i$.  Notice that the SM result can be easily recovered by taking
the limits $m_{\nu_j}= 0$ and $U^{ji} = \delta_{ji}$. Our result
translates into corrections to BR($W\to e \nu$), BR($W\to \mu\nu$) and
BR($W\to \tau\nu$), whose experimental values~\cite{Beringer:1900zz}
and SM predictions\footnote{Here we quote the 1-loop calculation
  of~\cite{Kniehl:2000rb}, where $m_H \sim 100$ GeV was
  used.}~\cite{Kniehl:2000rb} exhibit at present a small tension,
\begin{align}
\text{BR}(W\to e \nu)^\text{SM}&\,=\,0.108383&
\text{BR}(W\to e \nu)^\text{Exp}&\,=\,0.1080 \pm 0.009 \label{eq:Wellnu:e}\\
\text{BR}(W\to \mu \nu)^\text{SM}&\,=\,0.108383&
\text{BR}(W\to \mu \nu)^\text{Exp}&\,=\,0.1075 \pm 0.0013 \label{eq:Wellnu:mu} \\
\text{BR}(W\to \tau \nu)^\text{SM}&\,=\,0.108306&
\text{BR}(W\to \tau \nu)^\text{Exp}&\,=\,0.1057 \pm 0.0015\,. \label{eq:Wellnu:tau}
\end{align}
The tension between LEP-II results~\cite{Alcaraz:2006mx} and the SM
prediction on $W \to \ell \nu $ decays has not been given a large
attention but for a few exceptions, see for example~\cite{Filipuzzi:2012mg}.

\subsection{$\pmb{\tau}$ decays}\label{sec:Rtau}
Due to its comparatively large mass, the tau lepton can have both
leptonic and mesonic decays, for example into pions or kaons. 
Here we discuss how the corrections to the $W \ell \nu$ vertex can
affect the $W$-mediated tree-level $\tau$ decays.

\subsubsection*{Leptonic $\pmb{\tau}$ decays}
Flavour universality in leptonic $\tau$ decays is parametrised by the quantity 
$R_\tau$, 
\begin{equation}\label{eq:Rtau}
R_\tau \,\equiv \,
\frac{\Gamma(\tau^-\to \mu^-\nu\nu)}
{\Gamma(\tau^-\to e^-\nu\nu)}\,. 
\end{equation}
In the SM (with vanishing $m_\nu$) one has $R_\tau \simeq
0.973$~\cite{Pich:2009zza}. Experimentally, this observable  
has been measured to a precision better than
the individual decay widths by the BaBar~\cite{Aubert:2009qj} 
and CLEO~\cite{Anastassov:1996tc} experiments, with the following 
values 
\begin{equation}\label{eq:Rtau:exp}
R_\tau \,= \,0.9796\pm0.0016\pm0.0036 \,\, (\text{BaBar})\,,
\quad \quad
R_\tau \,= \, 0.9777\pm0.0063\pm0.0087\,\,
(\text{CLEO})\,, 
\end{equation}
while the current global 
fit stands at $R_\tau = 0.9764 \pm 0.0030$~\cite{Beringer:1900zz}. 

In the presence of additional sterile fermions, the decay width
$\Gamma(\ell_i \rightarrow \ell_j \nu \nu)$ must be corrected\footnote{New
  corrections to the SM results for the leptonic decay width of
  muons and taus have been recently discussed
  in~\cite{Ferroglia:2013dga,Fael:2013pja}, in the limit of massless neutrinos.}
and after summing over all the kinematically accessible neutrinos, one
finds, under the assumption of a Majorana nature for the neutrinos,
\begin{align}\label{eq:gammatauellnunu:maj}
 \Gamma_{\mathrm{tot}} &= \sum_{\alpha=1}^{N_\text{max}^{(\ell_j)}} \,
\sum_{\beta=1}^{\alpha} \Gamma_{\alpha\beta}\,,\\
 \mathrm{with} & \nonumber\\
 \Gamma_{\alpha\beta} &=\, \frac{G_F^2 \,
   (2-\delta_{\alpha\beta})}{m_{\ell_i}^3 \,(2\pi)^3}
 \int_{(m_{\ell_j}+m_{\nu_\alpha})^2}^{(m_{\ell_i}-m_{\nu_\beta})^2}
 \mathrm{d}s_{j\alpha} \left[\frac{1}{4} \,|U_{i\alpha}|^2\,
   |U_{j\beta}|^2 (s_{j\alpha} -m_{\ell_j}^2-m_{\nu_\alpha}^2)
  \, (m_{\ell_i}^2+m_{\nu_\beta}^2-s_{j\alpha}) \right.\nonumber\\ 
& \left. \quad \quad \quad + \frac{1}{2} \,\Re(U_{i\alpha}^* \,
   U_{j\beta} \,U_{i\beta} \,U_{j\alpha}^*) \,m_{\nu_\alpha}
   \,m_{\nu_\beta} \,
   \left(s_{j\alpha}-\frac{m_{\nu_\alpha}^2+m_{\nu_\beta}^2}{2}\right)
   \right]\nonumber\\ 
& \quad \quad \quad \times \frac{1}{s_{j\alpha}}\,
 \sqrt{(s_{j\alpha}-m_{\ell_j}^2-m_{\nu_\alpha}^2)^2-4m_{\nu_\alpha}^2m_{\ell_j}^2}
\, \sqrt{(m_{\ell_i}^2+m_{\nu_\beta}^2-s_{j\alpha})^2-
4\,m_{\nu_\beta}^2 \,m_{\ell_i}^2}\nonumber\\  
& + \alpha \leftrightarrow \beta\, ,
\end{align}
where $U$ is the full leptonic mixing matrix defined in Eq.~(\ref{eq:cc-lag})
and 
$G_F$ is the Fermi constant. The Dalitz variable is defined as
${s_{j\alpha}}=(p_{\ell_j}+p_{\nu_\alpha})^2$, $p_{\ell_j}$, $p_{\nu_\alpha}$ being the corresponding momenta for $\ell_j$ and $\nu_\alpha$.

\subsection*{Mesonic $\pmb{\tau}$ decays}
It is also interesting to consider the impact of the modified
$W\ell\nu$ vertex on mesonic $\tau$ decays.  In particular, we
consider the following observables
\begin{equation}\label{eq:GammaTauKnu:def}
R^{\ell \tau}_{K} \,\equiv\, 
\frac{\Gamma(\tau\to K \nu)}{\Gamma(K \to \ell \nu)}\, 
\quad \quad \text{and }\quad \quad 
R^{\ell \tau}_{\pi} \,\equiv\, 
\frac{\Gamma(\tau\to \pi \nu)}{\Gamma(\pi \to \ell \nu)}\,, 
\end{equation}
with $\ell=e\,, \mu$. These observables
allow to indirectly probe the universality of the 
$\tau$-coupling, while remaining 
free of hadronic matrix element uncertainties.
Indeed, the dependence on the decay constants cancels out
in these ratios at tree-level. The corresponding experimental values
can be computed from the individual decay widths given
in~\cite{Beringer:1900zz} 
\begin{eqnarray}\label{eq:GammaTauKnu:exp}
R^{\mu \tau}_{K} \,=\, 469.3\pm7.0\, 
&\text{and }
&R^{\mu \tau}_{\pi} \,=\, 9703 \pm 34\,,\\
R^{e \tau}_{K} \,=\, (1.886\pm0.078)\times 10^{7}\, 
&\text{and }
&R^{e \tau}_{\pi} \,=\, (7.888 \pm 0.038) \times 10^{7}\,,
\end{eqnarray}
while our estimations of the tree-level SM predictions are
\begin{eqnarray}\label{eq:GammaTauKnu:SM}
R^{\mu \tau}_{K} \,=\, 476.0\,
&\text{and } 
&R^{\mu \tau}_{\pi} \,=\, 9756\,,\\
R^{e \tau}_{K} \,=\, 1.853\times 10^7\,
&\text{and } 
&R^{e \tau}_{\pi} \,=\, 7.602 \times 10^7\,.
\end{eqnarray}
In the SM extended by the new sterile states, 
the mesonic $\tau$ decay width is given by 
\begin{eqnarray}\label{eq:GammaTauKnu}
\Gamma(\tau \to P \nu_i)\,=\,
\frac{G_F^2 \, f_P^2}{16 \pi\, m^3_\tau} \, 
| U^{3i}|^2\,
|V_\text{CKM}^{q q^\prime}|^2  \lambda^{1/2}(m_\tau,m_{P} ,m_{\nu_i})  
\left[(m_{\tau}^2- m_{\nu_i}^2)^2 - 
m_P^2 (m_{\nu_i}^2+m_{\tau}^2)  \right]\ , 
\end{eqnarray}
where $P=\pi, K$ and $i =1, \dots, N_\text{max}^{(P)}$. The function
$\lambda(m_\tau,m_{P} ,m_{\nu_i})$ has been given in
Eq.~(\ref{lambda.function})  and
$V_\text{CKM}^{q q^\prime}$ denotes the 
appropriate CKM matrix element. In addition,
$N_\text{max}^{(P)}$ is the heaviest neutrino mass
eigenstate which is kinematically allowed when a $P$ meson is produced. 
The leptonic pseudoscalar meson decay width in the presence of
additional sterile neutrinos was given in~\cite{Abada:2012mc},  and
will be discussed in the following
subsection (see Eqs.~(\ref{eq:RK:Rpi}, \ref{eq:RPresult})).

\subsection{Leptonic pseudoscalar meson decays}\label{sec:lepton.meson.decays}
We now address the decays of pseudoscalar mesons into leptons,
whose dominant contributions arise from tree-level $W$ mediated
exchanges. The theoretical prediction of some decays can be 
plagued by hadronic matrix element 
uncertainties (as is the case of leptonic $B$ decays):
however, by considering
the ratios 
\begin{equation}\label{eq:RK:Rpi}
R_P \, \equiv \,\frac{\Gamma (P^+ \to e^+ \nu)}{\Gamma (P^+ \to \mu^+
  \nu)}\,,
\end{equation}
these can be significantly reduced since the hadronic uncertainties cancel
out to a good approximation, so that the SM predictions can be
computed with a high precision. 
In order to compare the experimental bounds 
({some of which have recently been
obtained with an impressive precision}) with the SM expectation, it
proves convenient to use the quantity  $\Delta r_P$, which
parametrises deviation from the SM prediction, possibly arising from
new physics contributions: 
\begin{equation}\label{eq:deltar:P}
R_P \, = \,R_P^\text{SM} \, (1+\Delta r_P) \quad
\text{or equivalently}
\quad
\Delta r_P \, \equiv \, \frac{R_P}{R_P^\text{SM}} - 1\,.
\end{equation}

\noindent The expression for $R_P$ in the SM extended by sterile neutrinos
is given by~\cite{Abada:2012mc}
\begin{equation}\label{eq:RPresult}
R_P \,= \,\frac{\sum_i 
F^{i1} G^{i1}}{\sum_k F^{k2} G^{k2}}\,, \quad \text{with}
\end{equation}
\begin{align}
& F^{ij}\,=\, |U^{ji}|^2
\quad \text{and} \ \
 G^{ij} \,=\, \left[m_P^2 (m_{\nu_i}^2+m_{\ell_j}^2) - 
 (m_{\nu_i}^2-m_{\ell_j}^2)^2 \right] \, \lambda^{1/2}(m_{P} ,m_{\nu_i}, m_{\ell_j}), 
  \label{eq:FG}
\end{align}
where the function $\lambda(m_{P} ,m_{\nu_i}, m_{\ell_j})$ is given by
Eq.~(\ref{lambda.function}). Again, we recall that all states do not
necessarily contribute to $R_P$; this can be confirmed from inspection
of $G^{ij}$, which must be a positive definite quantity.

The result of Eq.~(\ref{eq:RPresult}) allows for a straightforward
interpretation of the impact of the new sterile states: $F^{ij}$
represents the impact of new interactions (absent in the SM), whereas
$G^{ij}$ encodes the mass-dependent factors.  In the limit where $m_{\nu_i}=
0$ and $U^{ji} = \delta_{ji}$, one can recover the SM result from
Eq.~(\ref{eq:RPresult}),
\begin{equation} \label{eq:RMSM}
R_P^\text{SM} = \frac{m_e^2}{m_\mu^2}
\frac{(m_P^2-m_e^2)^2}{(m_P^2-m_\mu^2)^2} \,, 
\end{equation}
to which small electromagnetic corrections (accounting for internal
bremsstrahlung and structure-dependent effects) should be
added~\cite{Cirigliano:2007xi}. Notice the strong helicity
suppression, 
$R_P^\text{SM} \propto \frac{m_e^2}{m_\mu^2}$ in Eq.\eqref{eq:RMSM}. This
makes $R_P$ (and $\Delta r_P$) one of the most sensitive observables to study
lepton flavour universality violation.

The general expression for $\Delta r_P$ reads
\begin{equation}\label{eq:deltaRPresult}
\Delta r_P \,= \,\frac{m_\mu^2 (m_P^2 - m_\mu^2)^2}{m_e^2 (m_P^2 - m_e^2)^2}\,
\frac{\operatornamewithlimits{\sum}_{m=1}^{N_\text{max}^{(e)}} 
F^{m1}\, G^{m1}}
{\operatornamewithlimits{\sum}_{n=1}^{N_\text{max}^{(\mu)}} 
F^{n2}\, G^{n2}} -1 \,,
\end{equation}
where $N_\text{max}^{(\ell_j)}$ is the heaviest neutrino mass
eigenstate kinematically allowed in association with $\ell_j$.  As can
be seen from the above equation, $\Delta r_P$ can considerably deviate
from zero, due to the mass hierarchy of the new states and the
active-sterile mixings. Owing to its analytical transparence, we again
stress the distinct sources of enhancement to $\Delta r_P$.  Firstly,
and if the new sterile states are light (in particular lighter than
the decaying meson) all the $\nu_i$ mass eigenstates can be
kinematically accessible as final states.  Although in this limit
unitarity would be recovered (as one would sum over all $3+N_s$ states
whose mixing is parametrised by $U$), $\Delta r_P$ can still be
enhanced due to the new phase space factors, see
Eq.~(\ref{eq:FG}). Heavier steriles can also lead to an enhancement of
$\Delta r_P$, as a result of deviations from unitarity: even though this is
more model-dependent, sterile mixings to active neutrinos can be
sizable due to the possibility of having larger Yukawa couplings. 

Although the general expression for the leptonic pseudoscalar decays
has been given above, Eqs.~(\ref{eq:RPresult},
\ref{eq:deltaRPresult}), we briefly comment below on each of the
specific observables we will address.

\subsubsection*{Light mesons: $\pmb{R_{K, \pi}}$ and $\pmb{R_{e,\mu}}$}
The $R_{K, \pi}$ (and $\Delta r_{K, \pi}$) observables were already 
discussed in \cite{Abada:2012mc}, and
constitute a perfect test of lepton flavour universality.
The comparison of theoretical
analyses~\cite{Cirigliano:2007xi,Finkemeier:1995gi} with the recent
measurements from the NA62 collaboration~\cite{Goudzovski:2011tc,Lazzeroni:2012cx} and
with the existing measurements on pion leptonic
decays~\cite{Czapek:1993kc}
\begin{eqnarray}\label{eq:RK:Rpi:SMvsexp}
& R_K^\text{SM} \, = \, (2.477 \pm 0.001) \, \times 10^{-5}\,,
\quad \quad
& R_K^\text{Exp} \, = \,(2.488 \pm 0.010) \, \times 10^{-5}\,,
\\
& R_\pi^\text{SM} \, =\, (1.2354 \pm 0.0002) \, \times 10^{-4}\,, 
\quad \quad
& R_\pi^\text{Exp} \, =\, (1.230 \pm 0.004) \, \times 10^{-4}\, ,
\end{eqnarray}
suggests that observation agrees at the $1 \sigma$ level with 
the SM predictions for 
\begin{equation}\label{eq:deltar:P:value} 
\Delta r_K \, = \, (4 \pm 4 )\, \times\, 10^{-3}\,, 
\quad \quad
 \Delta r_\pi \, = \, (-4 \pm 3 )\, \times\, 10^{-3}\,. 
\end{equation}
The current experimental uncertainty
in $\Delta r_K$ (of around 0.4\%) should be further reduced in the
near future, as one expects to have $\delta R_K / R_K \sim
0.1\%$~\cite{Goudzovski:2012gh,Strauch}, which can translate into 
measuring deviations $\Delta r_K \, \sim
\mathcal{O}(10^{-3})$.
There are also plans for
a more precise determination of $\Delta
r_\pi$~\cite{Pocanic:2012gt,Malbrunot:2012zz}.

\bigskip
We also consider the observables
\begin{equation}\label{eq:Re:Rmu}
R_e \, \equiv \,\frac{\Gamma (\pi^+ \to e^+ \nu)}{\Gamma (K^+ \to e^+
  \nu)}\quad , \quad R_\mu \, \equiv \,\frac{\Gamma (\pi^+ \to \mu^+
  \nu)}{\Gamma (K^+ \to \mu^+ \nu)}\ ,
\end{equation}
as well as the corresponding deviations from the SM predictions
\begin{equation}\label{eq:deltar:emu}
R_{e,\mu} \, = \,R_{e,\mu}^\text{SM} \, (1+\Delta
r_{e,\mu}) \quad \text{or equivalently} \quad \Delta r_{e,\mu} \,
\equiv \, \frac{R_{e,\mu}}{R_{e,\mu}^\text{SM}} - 1\,.
\end{equation}
Although at first sight apparently redundant, the study of the
observables $R_{e,\mu}$ and $\Delta r_{e,\mu}$ is well
motivated. Indeed, they  offer the possibility to extract the
ratios $f_\pi/f_K$ and $|V_{us}|/|V_{ud}|$ using an experimentally
clean signal and with little theoretical uncertainty. Unfortunately,
the current values of $R_e$ and $R_\mu$ are at present computed using
different measurements~\cite{Beringer:1900zz} coming from experiments
that are sometimes separated by more than 20 years. This makes a 
proper evaluation of systematic uncertainties quite
difficult. However, the NA62 experiment will have a good control
of the systematics due to the presence of the same charged lepton in
the final state (owing to the acquisition of both samples 
in the same data taking, with the same beam configuration and
with the same trigger strategy). In fact, $R_e$ can be measured with a
precision at the level of $0.5\%$ within a few years at
NA62~\cite{spadaro}. Even if  the experimental prospects for $R_\mu$ 
are less appealing, we nevertheless include it for completeness in our study.
The corresponding current experimental values are~\cite{Beringer:1900zz}
\begin{equation}
R_e = 3.70 \pm 0.02 \quad , \quad R_\mu = 0.748 \pm 0.002 \, ,
\end{equation}
and these can be combined with our estimates of the
SM results (central values), 
\begin{equation}
R_e^\text{SM} = 3.71258 \quad , \quad R_\mu^\text{SM} = 0.743103\,,
\end{equation}
leading to
\begin{equation}
\Delta r_e = -0.003 \pm 0.006 \quad , \quad \Delta r_\mu = 0.007 \pm 0.002 \, .
\end{equation}

\subsubsection*{Charmed mesons $\pmb{D, D_s}$: $\pmb{R_{D, D_s}}$ and $\pmb{R^{D}_{D_s}}$}
Nominal SM expectations on $D$ and $D_S$ leptonic decays, as well as
the experimental results from CLEO-c and BES III, can be found
in~\cite{Li:2011nij}.  In our study we focus in particular on $D_s$
observables, which have recently been well measured by
{CLEO-c}~\cite{Naik:2009tk}: 
\bea
\text{BR}(D_s\to\tau^+\nu)&=&(5.52\pm 0.57\pm 0.21)\times 10^{-2}\,,
\nonumber\\ 
\text{BR}(D_s\to\mu^+\nu)&=&(0.576\pm 0.045\pm0.054)\times 10^{-2}\,.  
\eea 
Both these observables present a
deviation from the theoretical expectation: 2.4 $\sigma$ using the QCD
sum rules estimation for the decay constant $f_{D_s}$, or a 2.8
$\sigma$ deviation using the lattice determination
(see~\cite{Naik:2009tk} and references therein).

As before, considering ratios of decay widths ($R_{D_{s}}$) allows to
cancel the theoretical uncertainties (due to $f_{D_s}$), and thus to
compute the SM prediction to a high precision,
\begin{equation}
\label{eq:RDs}
R_{D_{s}} \, \equiv \,\frac{\Gamma (D_s \to \tau \nu)}{ \Gamma ({D_{s} \to \mu
  \nu)}}\ .
\end{equation}

\medskip
Motivated by the fact that the ratio of $D_s$ and $D$ decay constants
was recently determined to a high precision~\cite{Becirevic:2013mp},
we have also studied the following ratio \bea
\label{eq:RDs_d}
R_{D_s^D} \, \equiv \, \frac{\Gamma (D_s \to \tau \nu)}{\Gamma (D \to \mu
  \nu)}\,\propto \,\frac {1}{ \lambda^2} 
\,\left| {\frac {f_{D_s}}{ f_{D}} }\right|^2\,, 
   \eea
where $\lambda$ is the CKM parameter in the Wolfenstein
parametrisation ($V_\text{CKM}^{cs}/V_\text{CKM}^{cd}=1/\lambda$) and $f_{D_s}$, $ f_D$
are the $D_s$ and $D$ pseudoscalar decay constants, respectively.   
Lattice QCD computations have allowed to determine the ratio 
$ {f_{D_s}}/{ f_D}$ with a high precision. For instance in~\cite{Aoki:2013ldr}, one has 
\bea
\frac {f_{D_s}}{ f_D}\,=\,  
1.187\pm 0.012\ ,
\label{eq:ratiofDsDflag}
\eea
while in~\cite{Becirevic:2013mp}, one has
\bea
\frac {f_{D_s}}{ f_D}\,=\,  
0.995(6)(4)\times \frac {f_{K}}{ f_\pi}\,,
\label{eq:ratiofDsD}
\eea
$f_{K}$ and $ f_\pi$ being the kaon and pion  decay 
constants, whose ratio is given by (world average value, 
see~\cite{Antonelli:2009ws}):
\bea
\frac {f_{K}}{ f_\pi}\,= \,1.194(5)\,.
\label{eq:ratiofKPi}
\eea

\subsubsection*{Leptonic $\pmb{B}$ meson decays: BR($\pmb{B \to \tau \nu}$)}
Similarly, the leptonic decays of heavier mesons can also be affected
by changes in the $W \ell \nu$ vertex, in particular $B \to \tau
\nu$.  In the SM extended by the new sterile states, the decay rate
for $B\to \tau \nu_i$ is given by: 
\begin{eqnarray}\label{eq:GammaBTaunu}
\Gamma(B\to \tau \nu_i)\,=\,
\frac{G_F^2 \, f_B^2}{8 \pi\, m^3_B} \, 
| U^{3i}|^2\,
|V_\text{CKM}^{ub}|^2  \lambda^{1/2}(m_B,m_{\tau} ,m_{\nu_i})  
\left[m_B^2 (m_{\nu_i}^2+m_{\tau}^2) -
(m_{\tau}^2- m_{\nu_i}^2)^2  \right]\ ,
\end{eqnarray}
where $i =1, \dots, N_\text{max}$ ($N_\text{max}$ corresponding to the heaviest
$\nu_s$ which is kinematically allowed when the 
$B$ meson decays into a $\tau$ lepton). The function
$\lambda(m_{B} ,m_\tau,m_{\nu_i})$ has been given in
Eq.~(\ref{lambda.function}).

The following bounds must be 
taken into account~\cite{Beringer:1900zz}:
\begin{align}
 \text{BR}(B \rightarrow e \nu) &< 9.8 \times 10^{-7}\,,\\
 \text{BR}(B \rightarrow \mu \nu) &< 10^{-6}\,,\\
 \text{BR}(B \rightarrow \tau \nu) &= (1.65 \pm 0.34) \times 10^{-4}\,.
\end{align}
It should be noted that the experimental measurement of $\text{BR}(B
\to \tau \nu) $ significantly   
deviates from its SM prediction,
\begin{eqnarray}\label{eq:BTAUNU:SMvsexp}
 \text{BR}^\text{SM}(B \to \tau \nu) \, = 
\, ( 0.83(8)(6) ) \, \times 10^{-4}\,,
\end{eqnarray}
which was obtained with $V_\text{CKM}^{ub}=3.65(13)\times 10^{-3}$, corresponding to 
the average estimate from the global fits of 
CKMfitter~\cite{Charles:2011va} and UTfit~\cite{Bona:2009cj}, and
$f_B=(188\pm 6)$ MeV, an average from the most recent lattice QCD 
results~\cite{Dowdall:2013tga,Bazavov:2011aa,Blossier:2011dg}. 
Interestingly, the Belle collaboration has published an updated
analysis, reporting an even lower value~\cite{Adachi:2012mm},  
\begin{equation}
\text{BR}(B \to \tau \nu) \, = \, (0.72 ^{+0.27}_{-0.25}\pm
0.11) \, \times 10^{-4}\, ,
\end{equation}
and when averaged with the
BaBar result~\cite{Aubert:2007xj}, the Belle measurement leads to
\begin{eqnarray}\label{eq:BTAUNU:SMvsexp2}
\text{BR}(B \to \tau \nu) \, 
= \, (1.15\pm 0.23) \, \times 10^{-4}\, . 
\end{eqnarray}
Notice that since only the decay $B \to \tau \nu$ has been observed,
it is not possible to study a 
ratio of decay widths as done for other pseudocalar mesons.

\subsection{Semileptonic pseudoscalar meson decays}
Recent (surprising) experimental results for the ratio of 
the branching fractions of the 
$B \to D^{(*)}\tau \nu$ and $B \to D^{(*)}\mu \nu$ decays 
have opened the door to the possibility of constraining potential New
Physics contributions through these decay modes. 
In our analysis, we only focus on the following 
observable\footnote{We recall that the form factors for $B\to D^*\ell \nu$ 
are poorly known and have very large theoretical uncertainties.} 
\begin{equation}\label{eq:RDSL:def}
R(D)\, \equiv \,
\frac {\text{BR} (B^+ \to D \,\tau^+ \,\bar\nu_\tau)} 
{\text{BR} (B^+ \to D \,\mu^+\,
  \bar\nu_\mu) }\,,
\end{equation}
for which BaBar's recent measurement~\cite{Lees:2012xj} is  
\begin{equation}\label{eq:RDSL:exp}
R(D)^\text{Exp}= 0.440\pm0.058_\text{ stat.}\pm0.042_\text{ syst.}.
\end{equation}
Notice that BaBar's definition of $R(D)$ does not
distinguish a muon from an electron in the final state, 
i.e., the observable in fact corresponds to $R(D) =
\frac{\Gamma (B^+ \to D \,\tau^+ \nu)}{\Gamma (B^+ \to D\, \ell^+ \nu)}$.

Based on lattice estimations of the hadronic matrix elements
(parametrised by form factors), the SM 
prediction for $R(D)$ is~\cite{Becirevic:2012jf}
\begin{equation}\label{eq:RDSL:th}
R(D)^\text{SM}\,  = \,0.31\pm0.02\,,
\end{equation} 
which lies more than $1 \sigma$ (but less than $2 \sigma$) below  
the experimental results\footnote{This SM estimation is consistent with a
  different theoretical prediction made in~\cite{Fajfer:2012vx}.}.

It is worth mentioning that the BaBar excess of events in $B\to
D^*\,\tau\,\nu$ decays, revealing a 3.4 $\sigma$ deviation from the SM
prediction~\cite{Luth:2012ng}, cannot be accommodated by contributions
from charged Higgs bosons in the context of type-II Two Higgs Doublet Model
(2HDM)~\cite{Celis:2012dk}.  In fact, BaBar excludes the latter
contributions as an explanation of the deviations in $R(D)$ and
$R(D^*)$ from their SM predictions~\cite{schwanda-portoroz}.
Moreover, and as discussed in~\cite{Lees:2013uzd}, the observed excess
on $R(D)$ (and $R(D^*)$) also fails to be explained in large portions
of the more general type-III 2HDM. Similarly, the authors of
Ref.~\cite{Celis:2012dk} found regions in the parameter space of the
Aligned 2HDM able to accommodate the $R(D^*)$ measurement, although in
conflict with the constraints from leptonic charm decays.

For these observables, the mass of the neutral leptons is usually
neglected. Here we will derive analytical expressions for $\Gamma(P\,
\to P^\prime \, + \, \ell_i\, +\, \nu_j)$ in terms of
invariants, keeping all lepton masses (neutral and charged ones).
Notice that $R(D)$ is more interesting since tests of compatibility
of the Standard Model (or any of its extensions) can be done
experimentally with a {\it minimal theory input}. Indeed, the decay
rates are parametrised by two form factors, $F^+(q^2)$ and $F^0(q^2)$.
The first, $F^+(q^2)$, has  recently  been experimentally well measured~\cite{Aubert:2009ac} 
and the behaviour of the second,
$F^0(q^2)$, with respect to the transfer momentum $q^2$,  has also been
determined~\cite{Becirevic:2012jf} (consistent with many different
theoretical estimations from Lattice QCD
collaborations~\cite{Bailey:2012rr,deDivitiis:2007ui}, as well as with
QCD sum rules analyses~\cite{Faller:2008tr,Azizi:2008tt}),  
\bea
\frac{F^0(q^2)} {F^+(q^2)}= 1- \alpha\ q^2,\quad \alpha =
0.022(1)\ \text{GeV}^{-2}.  \eea Consider then the semileptonic meson
decay
\begin{equation}
P\, \to P^\prime \, + \, \ell_i\, +\, \nu_j \, ,
\end{equation}
with $m$ the mass of the decaying pseudoscalar meson, 
$m_{1,2}$ those of the final state charged
and neutral leptons, and $m_3$ the mass of the final pseudoscalar  state meson. 
The total width of the decay can be decomposed as 
\begin{equation}
\Gamma_\text{tot}\, = \, 
\Gamma_{c_1} \, + \,\Gamma_{c_2} \, + \,\Gamma_{c_3} \, + \,\Gamma_{c_4}\,,
\end{equation}
where each (partial) width is associated to the form factors 
$F^+ (q^2)$, $F^0 (q^2)$ (and combinations thereof) as follows
\begin{align}
\Gamma_{c_1, c_2}  \rightsquigarrow |F^+ (q^2)|^2\, ;
\quad 
\Gamma_{c_3}  \rightsquigarrow |F^0 (q^2)|^2\, ;
\quad 
\Gamma_{c_4} \rightsquigarrow 2 \operatorname{Re}(F^0 F^{+ *})\,.
\end{align}

The above widths can be written as:
\begin{align}
& \Gamma_{c_1}\, = \, 
\frac{G_F^2}{192 \pi^3}\,
\frac{|V_\text{CKM}^{q q^\prime}|^2\, |U_{ij}|^2}{m^3}
\,\int_{\left(m_1+m_2\right)^2}^{\left(m-m_3\right)^2} dq^2\, |F^+
(q^2)|^2\, \lambda^{{3/2}}(q,m,m_3)\, \lambda^{{3/2}}(q,m_1,m_2)
\,\frac{1}{q^6}\,, 
\nonumber \\
& \Gamma_{c_2}= 
\frac{G_F^2}{128 \pi^3}
\frac{|V_\text{CKM}^{q q^\prime}|^2\, |U_{ij}|^2}{m^3}
\,\int_{\left(m_1+m_2\right)^2}^{\left(m-m_3\right)^2} dq^2
|F^+ (q^2)|^2\,
\lambda^{{3/2}}(q,m,m_3)\, \lambda^{{1/2}}(q,m_1,m_2)
\,\frac{1}{q^6}\,\times \nonumber\\ 
&\hskip 6cm \left[
q^2(m_1^2 +m_2^2) - (m_1^2 - m_2^2)^2
\right ],
\nonumber \\
& \Gamma_{c_3}\, = \, 
\frac{G_F^2}{128 \pi^3}\,
\frac{|V_\text{CKM}^{q q^\prime}|^2\, |U_{ij}|^2}{m^3}
\,\int_{\left(m_1+m_2\right)^2}^{\left(m-m_3\right)^2}dq^2\,
|F^0 (q^2)|^2\, 
\left (
\frac{\Delta m^2}{q^2}
\right )^2 \,\lambda^{{1/2}}(q,m,m_3) \, \frac{1}{q^2}\, \times \nonumber\\
&\hskip 6cm\lambda^{{1/2}}(q,m_1,m_2)\,\left[
q^2(m_1^2 +m_2^2) - (m_1^2 - m_2^2)^2
\right ],
\nonumber \\
& \Gamma_{c_4}\, = \,0\,. 
\end{align}
In the above expressions,  $V_\text{CKM}^{q q^\prime}$ denotes the
appropriate CKM matrix element, $q$ the momentum transfer, $ \Delta
m^2$ refers to the squared mass difference between the two meson
masses 
\bea \Delta m^2  \, = \, m^2 -m_3^2\, , 
\eea
and the function $\lambda(q,m_1,m_2)$ is given by Eq.~(\ref{lambda.function}).

\section{Numerical results and discussion}\label{sec:results}

As mentioned before, the non-unitarity of the leptonic mixing matrix,
together with the possible phase space enhancement (when the sterile
states are very light), can modify the contributions to rates for
leptonic and hadronic processes with  neutrinos in the final state. 
Observables which have been measured with
very good precision, and are in agreement with SM expectations, will
allow to constrain the departure from $\sum_i |U^{ji}|^2 = 1$ (as
seen in Section~\ref{sec:lepton.meson.decays}, the observables depend on
$\sum_i |U^{ji}|^2$, with the sum extending over all kinematically
accessible neutrino states).

In our approach we assume that all NP effects lie in the lepton
sector; notice however, that the decorrelation of NP effects arising
from the 
modified $W \ell \nu$ vertex is sometimes hampered by large systematic
uncertainties on the hadronic matrix elements (with impact on $V_\text{CKM}$
element determination). Moreover, as already mentioned, we do not
address higher-order corrections (had we computed the latter, 
the systematic errors related to the uncertainty in
hadronic matrix elements would overlap with our own
systematics). 

Although the generic idea explored in this work applies to any model
where the active neutrinos have sizable mixings with some additional
singlet states, in order to evaluate the contributions of the new
states, one must consider a specific framework. Here, as an
illustrative example, we consider the case of the Inverse
Seesaw~\cite{Mohapatra:1986bd} to discuss the potential of a model
with sterile neutrinos regarding tree-level contributions to leptonic
and semileptonic meson decays (as mentioned before, there are other
possible frameworks to illustrate the effect of sterile neutrinos on these
observables).

\subsection{The inverse seesaw}
\label{Subsec:ISS}
In the inverse seesaw~\cite{Mohapatra:1986bd}, the SM particle content
is extended by $n_R$ generations of right-handed (RH) neutrinos
$\nu_R$ and $n_X$ generations of singlet fermions $X$ (such that
$n_R+n_X = N_s$), both with lepton number
$L=+1$~\cite{Mohapatra:1986bd}. Even if deviations from unitarity can
occur for different values of $n_R$ and $n_X$, here we will consider
the case $n_R = n_X = 3$.  The Lagrangian is given by
\begin{equation}
\label{eq:L_IS}
\mathcal{L}_\text{ISS} = 
\mathcal{L}_\text{SM} - Y^{\nu}_{ij}\, \bar{\nu}_{R i} \,\tilde{H}^\dagger  \,L_j 
- {M_R}_{ij} \, \bar{\nu}_{R i}\, X_j - 
\frac{1}{2} {\mu_X}_{ij} \,\bar{X}^c_i \,X_j + \, \text{h.c.}\,,
\end{equation}
where $i,j = 1,2,3$ are generation indices and $\tilde{H} = i \sigma_2
H^*$. Notice that the present lepton number assignment for the new
states, together with
$L=+1$ for the SM lepton doublet, implies that the Dirac-type
right-handed neutrino mass term $M_{R_{ij}}$ conserves lepton number,
while the Majorana mass term $\mu_{X_{ij}}$ violates it by two
units.

The non-trivial structure of the neutrino Yukawa couplings $Y^\nu$
implies that the left-handed neutrinos mix with the right-handed 
ones after electroweak symmetry breaking.  
In the $\{\nu_L,{\nu^c_R},X\}$ basis,
one has the following symmetric ($9\times9$) mass matrix
$\mathcal{M}$,
\begin{eqnarray}
{\cal M}&=&\left(
\begin{array}{ccc}
0 & m^{T}_D & 0 \\
m_D & 0 & M_R \\
0 & M^{T}_R & \mu_X \\
\end{array}\right) \, .
\label{nmssm-matrix}
\end{eqnarray}
Here $m_D= \frac{1}{\sqrt 2} Y^\nu v$,  $v/{\sqrt 2}$ being the vacuum expectation
value of the SM Higgs boson.  Assuming $\mu_X \ll m_D \ll M_R$, the
diagonalization of ${\cal M}$ leads to an effective Majorana mass
matrix for the active (light) neutrinos~\cite{GonzalezGarcia:1988rw},
\begin{equation}\label{eq:nu}
m_\nu \,\simeq \,{m_D^T\, M_R^{T}}^{-1} \,\mu_X \,M_R^{-1}\, m_D \, ,
\end{equation}
while the remaining 6 sterile states have masses approximately given
by $M_{\nu} \simeq M_R$.

In what follows, and without loss of generality, we work in a basis
where $M_R$ is a diagonal matrix (as are the charged lepton Yukawa
couplings). The couplings $Y^\nu$ can be written using a modified Casas-Ibarra
parametrisation~\cite{Casas:2001sr} (thus automatically complying with
light neutrino data):
\begin{equation}\label{eq:CI:param}
Y^\nu \,= \,\frac{\sqrt{2}}{v} \, D^\dagger \, \sqrt{\hat M} \, R \,
\sqrt{{\hat m}_\nu} \, U_\text{PMNS}^\dagger \, ,
\end{equation}
where $\sqrt{{\hat m}_\nu}$ is a diagonal matrix containing the square
roots of the three eigenvalues of $m_\nu$ (cf. Eq. \eqref{eq:nu});
likewise $\sqrt{\hat M}$ is a diagonal matrix with the square roots
of the eigenvalues of $M = M_R \mu_X^{-1} M_R^T$.  The matrix
$D$ diagonalizes $M$ as $D M D^T = \hat{M}$, 
and $R$ is a $3 \times 3$ complex
orthogonal matrix, parametrised by $3$ complex angles, encoding the
remaining degrees of freedom.

A distinctive feature of the ISS is that the additional $\mu_{X_{ij}}$
parameters allow to accommodate the smallness of the active neutrino
masses $m_\nu$ for a low seesaw scale, but still with natural Yukawa
couplings ($Y^\nu\sim {\mathcal{O}}(1) $).  As a consequence, one can
have sizable mixings between the active neutrinos and the additional
sterile states. This is in contrast with the canonical type-I seesaw,
where having $\mathcal{O}(1)$ Yukawa couplings usually requires 
$M_R \sim 10^{15}$ GeV, thus leading to truly negligible
active-sterile mixings. 

The neutrino mass matrix is diagonalized as $U^T_\nu \mathcal{M} U_\nu
= \text{diag}(m_i)$. The nine neutrino mass eigenstates enter the
leptonic charged current through their left-handed component (see
Eq.~(\ref{eq:cc-lag}), with $i = 1, \dots, 9$, $j = 1, \dots,
3$). Again in the basis where the charged lepton mass matrix is
diagonal, the leptonic mixing matrix $U$ is given by the the
rectangular $3 \times 9$ sub-matrix corresponding to the first three
columns of $U_\nu$.

Finally, we also refer to
\cite{Forero:2011pc,Malinsky:2009gw,Dev:2009aw} for earlier studies on
non-unitarity effects in the inverse seesaw.

\subsection{Constraining the ISS parameter space}\label{sec:scan}
The adapted Casas-Ibarra parametrisation for $Y^\nu$,
Eq.~(\ref{eq:CI:param}), ensures that neutrino oscillation data is
satisfied (we use the best-fit values of the global analysis
of~\cite{Tortola:2012te} - see Eqs.~(\ref{eq:neutrino.data.NH},
\ref{eq:neutrino.data.IH})). 
The $R$ matrix angles are
taken to be real, their value randomly varied in the range 
${\theta}_{i}\in [0,2 \pi]$
(thus no contributions to lepton electric dipole
moments are expected).
However, we have verified
that similar results are found when considering the
more general complex $R$ matrix case.
We also study both hierarchies for the light neutrino spectrum, 
and the effect of a non-vanishing Dirac CP violating phase.

The simulated points are then subject to all the constraints
previously mentioned, and those not complying with the different
bounds (with the exception of the cosmological constraints) are
excluded. As argued in~\cite{Gelmini:2008fq}, the cosmological bounds
can be evaded by considering a non-standard-cosmology; we therefore
keep these points, explicitly identifying them via a distinctive
colour scheme (blue points in agreement with cosmological bounds, red
points requiring a non-standard cosmology) throughout the numerical
analysis and subsequent discussion. For illustrative purposes, we also 
display points excluded by the recent MEG bound 
(see Eq.~\ref{eq:BR:muegamma:sterile})), although these will always 
be identified by a different colour scheme (grey points in this case). 

\bigskip
Before proceeding to the analysis of the observables, let us briefly
discuss the effect of the above mentioned constraints on the potential
deviations from unitarity of the $\tilde
U_\text{PMNS}$ matrix, which is parametrised by $\tilde
\eta$
\beq
\tilde \eta = 1 - |\text{Det}(\tilde U_\text{PMNS})| \, ,
\eeq
considering both cases of normal (NH) and inverted (IH) hierarchies for the light
neutrino spectrum.

\begin{figure}[h!]
\begin{center}
\begin{tabular}{cc} 
\hspace*{-7mm}
\psfig{file=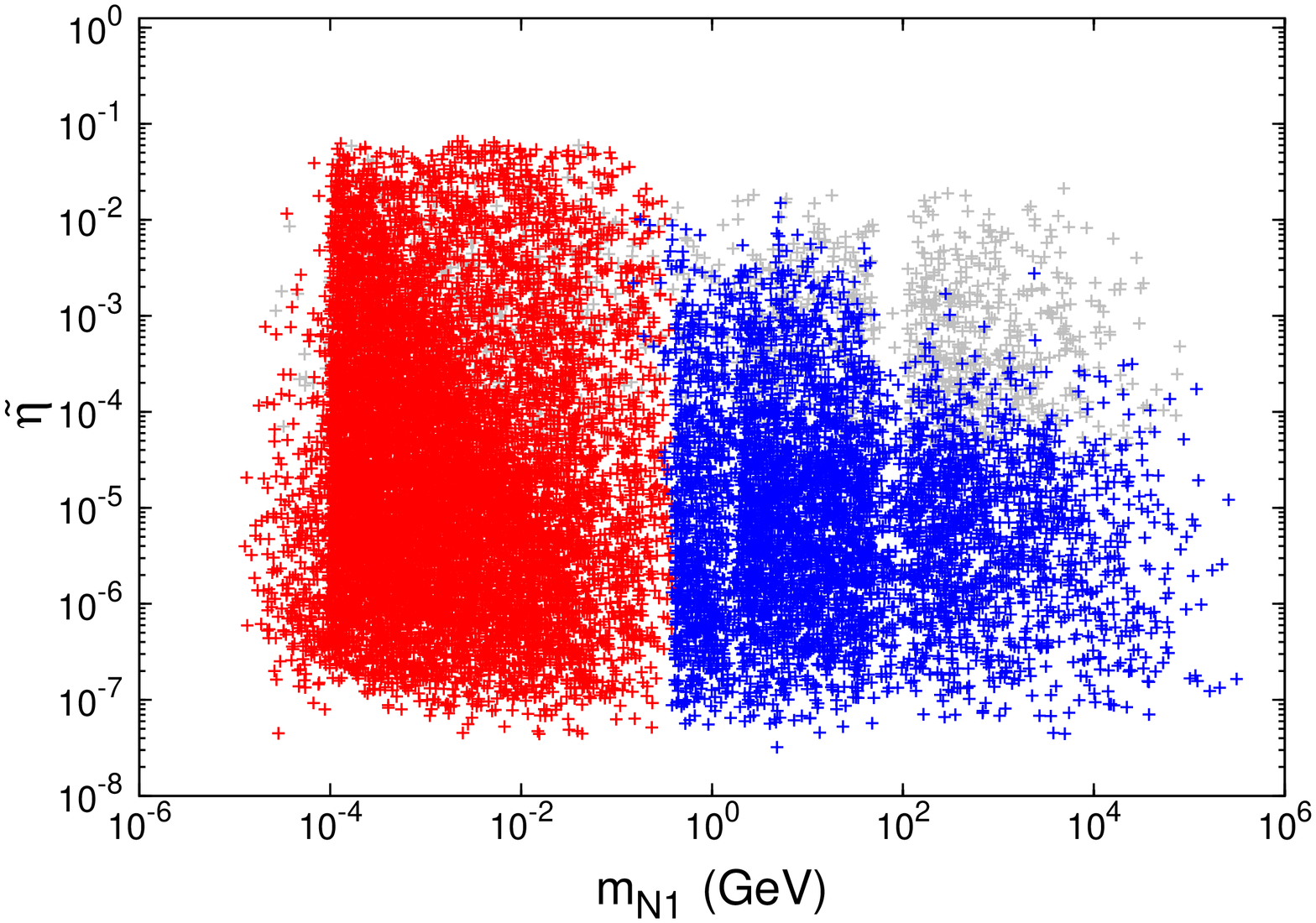, width=58mm,
  clip=, angle=270} 
&
\psfig{file=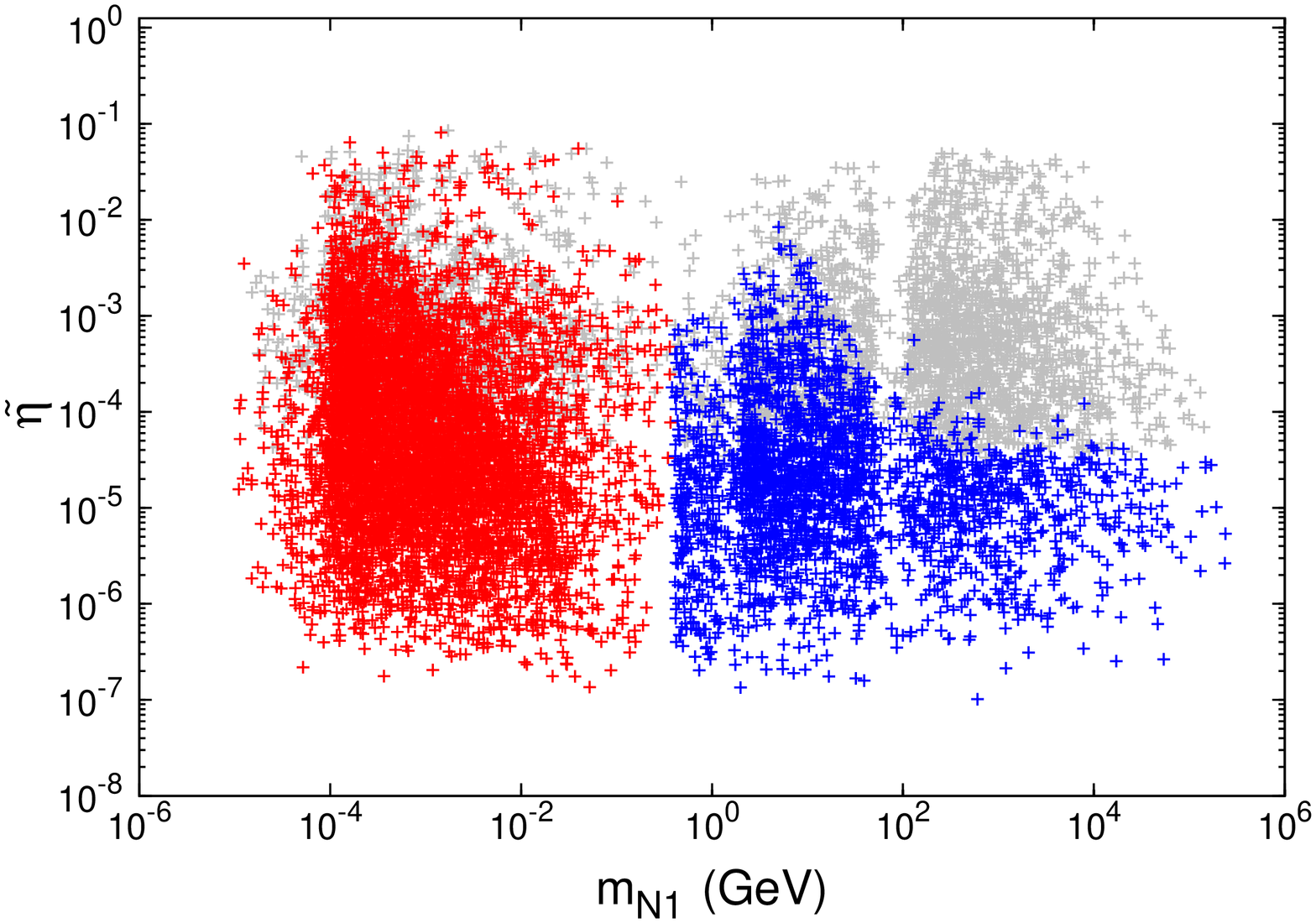, width=58mm,
  clip=, angle=270} 
\end{tabular}
\caption{Deviation from unitarity of the $\tilde
U_\text{PMNS}$ matrix, parametrised by $\tilde \eta$, as a function of the lightest
sterile neutrino mass, $m_{N_1}$, for normal (left) and inverted hierarchical
(right) light neutrino spectra. Blue points
are in agreement with cosmological bounds, while the red ones would require
considering a non-standard cosmology.  Grey points correspond to an associated 
BR($\mu \to e \gamma$) already excluded by MEG.} 
\label{fig:eta.mN1.H1H2}
\end{center}
\end{figure}

As can be seen from Fig.~\ref{fig:eta.mN1.H1H2}, regimes of $\tilde
\eta \sim \mathcal{O}(10^{-1})$ are indeed possible; however
these solutions are typically disfavoured from standard cosmology
arguments. It is nevertheless clear that the ISS framework favours a NH
scheme (notice that the density of points surviving the above
mentioned constraints is denser in this case). Moreover, the
recent MEG bound has a more severe impact in the case of IH scheme, 
excluding larger portions of the parameter space (here illustrated in
the $\tilde \eta - m_{N_1}$ plane). 
Notice that values $\tilde \eta \gtrsim
\mathcal{O}(10^{-1})$ are excluded since in this limit 
the seesaw condition is not satisfied.

\begin{figure}[h!]
\begin{center}
\psfig{file=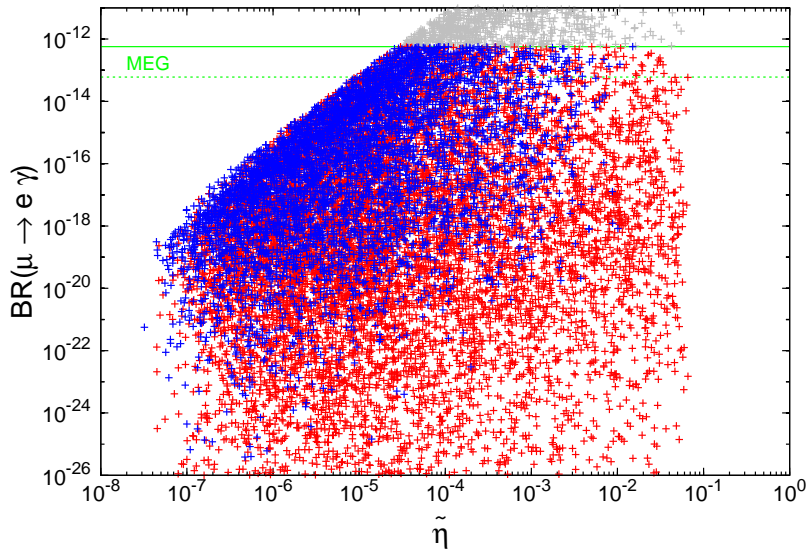,
  width=80mm, clip=, angle=270} 
\caption{BR($\mu \to e \gamma$) as a function of $\tilde \eta$, for
  the case of a NH spectrum.  Colour code as in Fig.~\ref{fig:eta.mN1.H1H2}.  Green
  horizontal lines denote current experimental bounds (solid) and MEG
  future sensitivity~\cite{Adam:2013mnn} (dashed).
}\label{fig:BRmuegamma.eta}
\end{center}
\end{figure}

The prospects concerning the observation of BR($\mu \to e
\gamma$) at MEG, as well as the impact of the current bound 
regarding the parameter space surveyed in our analysis, 
are collected in Fig.~\ref{fig:BRmuegamma.eta}. 
Given the significant constraints
on the parameter space arising from this observable, it would be 
undoubtably interesting to consider the actual impact of 
BR$(\mu \to e \gamma )$ on other parameters of the model, such as 
${\mu_X}_{ij}$, or the active-sterile mixing angles
$\theta_{i\alpha}$. However, and as can be inferred from the
description of the underlying numerical scan,
the fact that one has 
explored all degrees of freedom of the Yukawa couplings 
precludes this (for example, the impact of a given 
texture and/or regime for ${\mu_X}$
and $M_R$ would be mitigated by the mixings introduced via the
$R$-matrix). In other words, it is not possible to 
individually explore the effect of a given parameter on  
observables, as the latter simultaneously depend on a number 
of (varying) parameters.
Likewise, considering the dependence of an observable
on a given active-sterile mixing angle $\theta_{i\alpha}$ would
provide little insight, as various 
physical mixing regimes arise from the scan.

In what follows, we will consider 
a NH light neutrino spectrum in our analysis of the different
observables. Unless otherwise stated, the Dirac CP violating phase 
$\delta$ will be set to zero.

\subsection{Invisible $\pmb{Z}$ decays and $\pmb{W \to \ell \nu}$ }
We begin our analysis by discussing the potential contributions of the
additional sterile states to EW observables, in particular the
invisible $Z$ decay width (which is an EW precision test) and leptonic $W$
decays. 

\begin{figure}[h!]
\begin{center}
\begin{tabular}{cc} 
\hspace*{-7mm}
\psfig{file=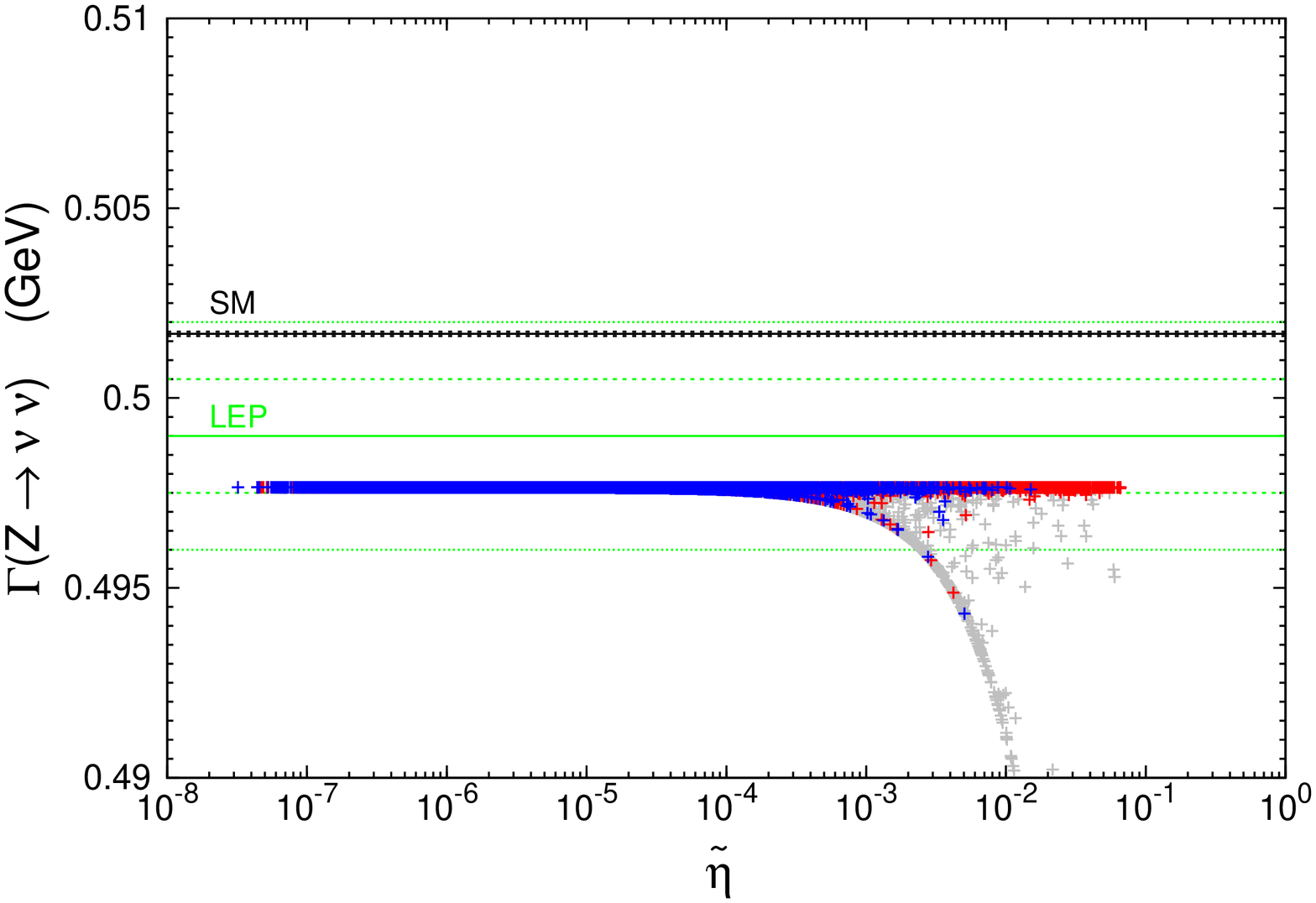, width=58mm,
  clip=, angle=270} 
&
\psfig{file=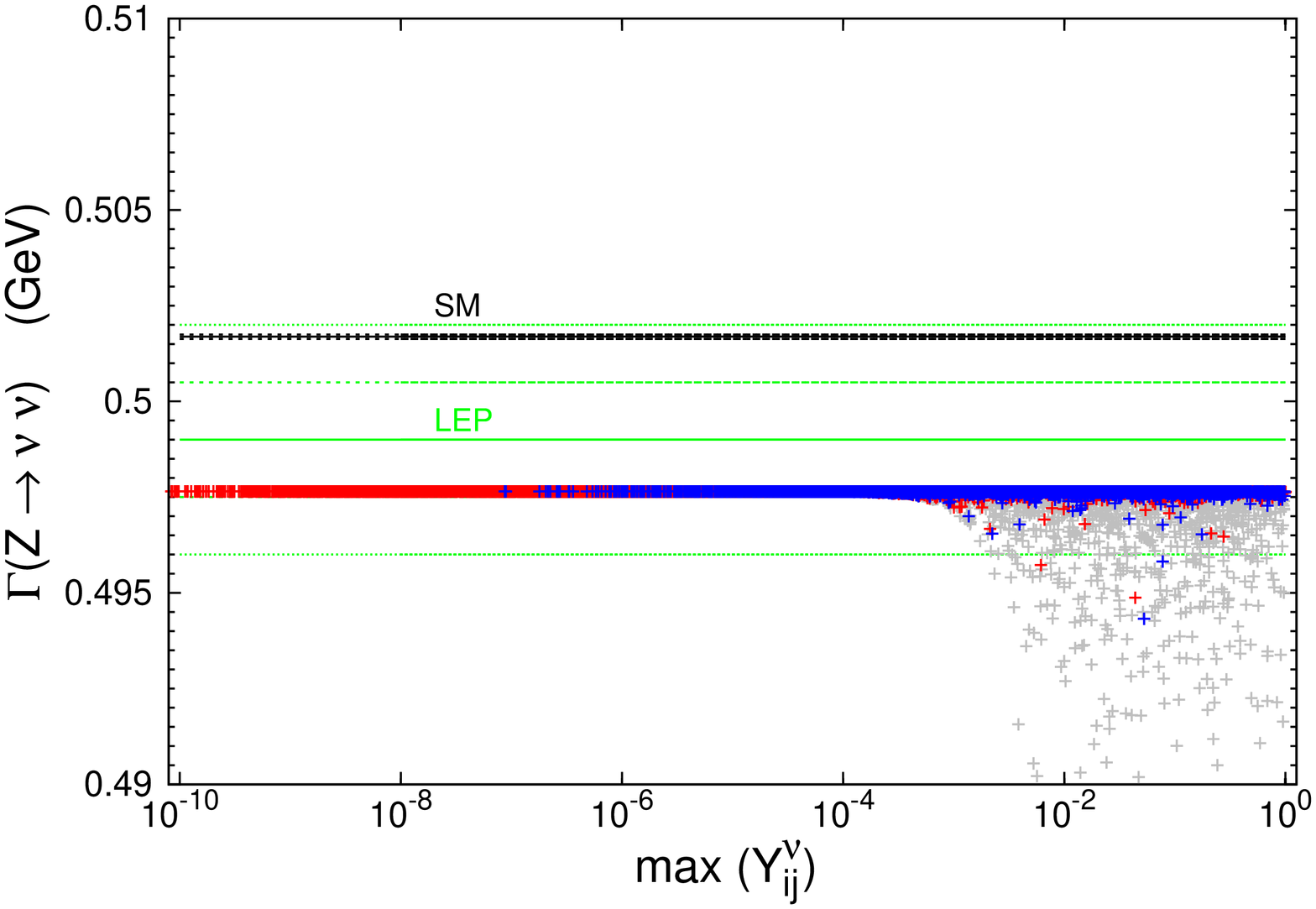, width=58mm,
  clip=, angle=270} 
\end{tabular}
\caption{$\Gamma (Z \to \nu \nu)$ as a function of $\tilde \eta$ (on
  the left) and of  
the maximal entry of $Y^\nu_{ij} / 4\pi$ (on the right). 
Blue points
are in agreement with cosmological bounds, while the red ones would require
considering a non-standard cosmology. Black lines
denote the SM prediction, green ones correspond to the experimental
measurement (full/dashed/dotted corresponding to central value,
$1 \sigma$ and $2 \sigma$ intervals, respectively). Grey points correspond to an associated 
BR($\mu \to e \gamma$) already excluded by MEG.} 
\label{fig:gammaznunu:eta}
\end{center}
\end{figure}

On the left panel of Fig.~\ref{fig:gammaznunu:eta}, we display the
invisible $Z$ decay width, $\Gamma(Z \to \nu \nu)$ (see
Eqs.~(\ref{eq:Znunu:sum}-\ref{eq:Znunu:couplings})), as a function of
$\tilde \eta$.
A black horizontal line denotes the SM
prediction (cf. Eq.~(\ref{eq:Znunu:SM})), while green lines
(full/dashed/dotted) correspond to the experimental measurements from
LEP (central, $1 \sigma$ and $2 \sigma$ intervals, respectively, see
Eq.~(\ref{eq:Znunu:Exp})).  
The apparent difference between the SM line (corresponding to a full
computation of this observable) and our 
SM-limit (obtained for the regime of very small $\tilde \eta$, for
which the PMNS becomes unitary) in Fig.~\ref{fig:gammaznunu:eta}  
is due to the fact that the latter is based on a tree-level
computation; should higher order corrections be taken into account,  
all points would be shifted towards the SM theoretical prediction.

Clearly, and even
though the non-unitarity of the PMNS only indirectly affects $Z \to
\nu \nu$ decays (via the small sterile component in the active light
neutrino eigenstates) this EW precision observable is a crucial consistency 
check of any model with extra sterile fermions, as the invisible decay width 
cannot exceed the SM prediction~\cite{Akhmedov:2013hec,Basso:2013jka}.
As can be seen in the left  panel of Fig.~\ref{fig:gammaznunu:eta}, 
a sizable reduction 
of the invisible decay width could indeed occur for  a regime of large $\tilde \eta$.
However,  this reduction of $\Gamma(Z \to \nu \nu)$ is precluded by the 
current MEG bound on BR($\mu \to e \gamma$).

These regimes of large $\tilde \eta$ are associated with large values
of the Yukawa couplings $Y^\nu_{ij}$: 
having such large  values of $Y^\nu_{ij}$ for a comparatively low
seesaw scale is a direct consequence of 
an inverse seesaw as the underlying framework for sterile
neutrinos (this effect was already 
discussed in~\cite{Abada:2012mc}, in relation to LFU violation in light
pseudoscalar meson decays). For $Y^\nu\sim \text{few} \times 10^{-2}$,
large active-sterile mixings can occur, possibly leading to a decrease
in the $Z$ boson decay width (in agreement
with~\cite{Akhmedov:2013hec,Basso:2013jka}).  
This can be confirmed on the right panel of 
Fig.~\ref{fig:gammaznunu:eta}, where we display $\Gamma(Z \to \nu \nu)$
versus the corresponding largest entry of the neutrino Yukawa
couplings, max($Y^\nu_{ij}$). 

\begin{figure}[h!]
\begin{center}
\begin{tabular}{cc}
\hspace*{-7mm}
\psfig{file=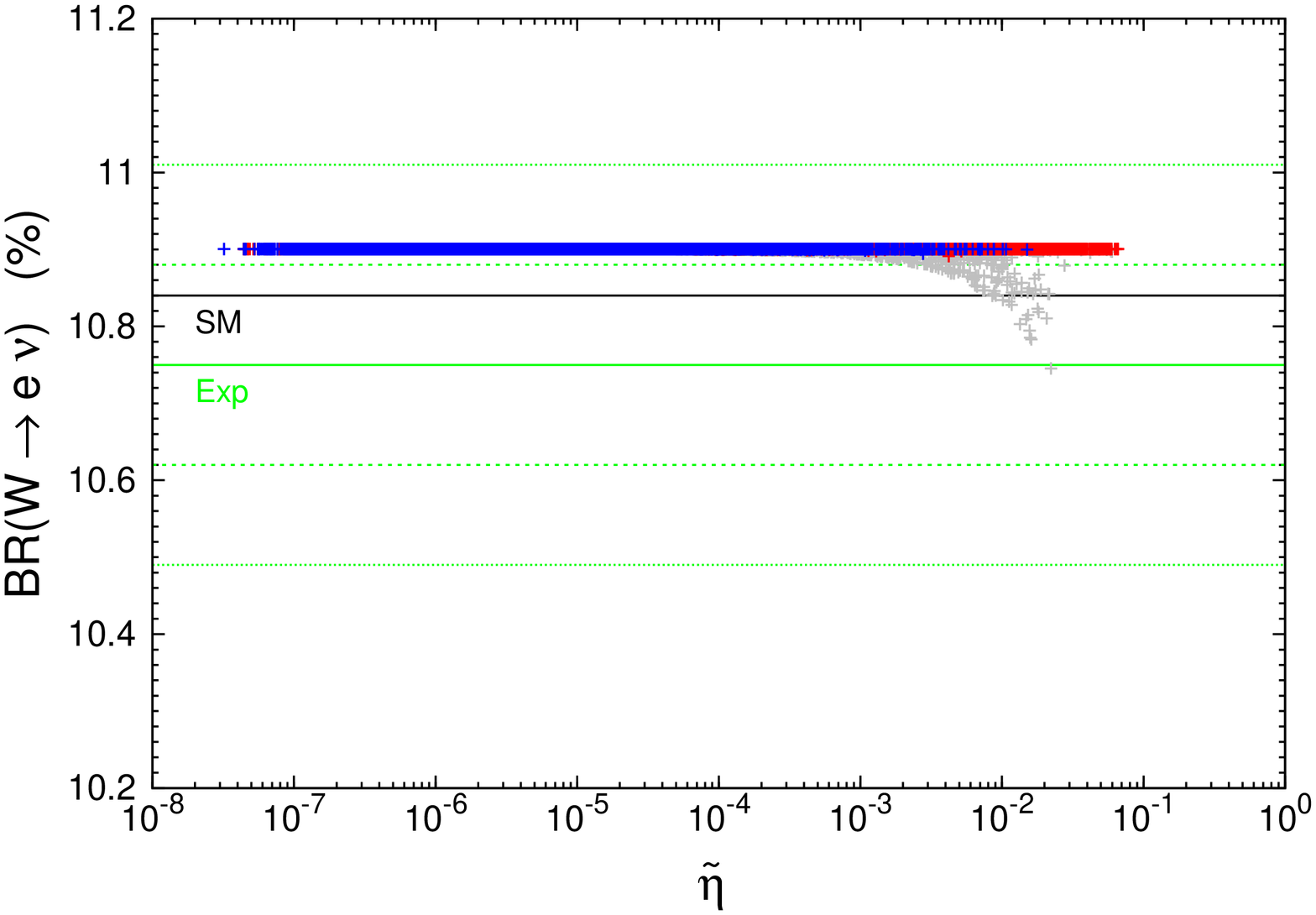, width=58mm,
  clip=, angle=270} & 
\psfig{file=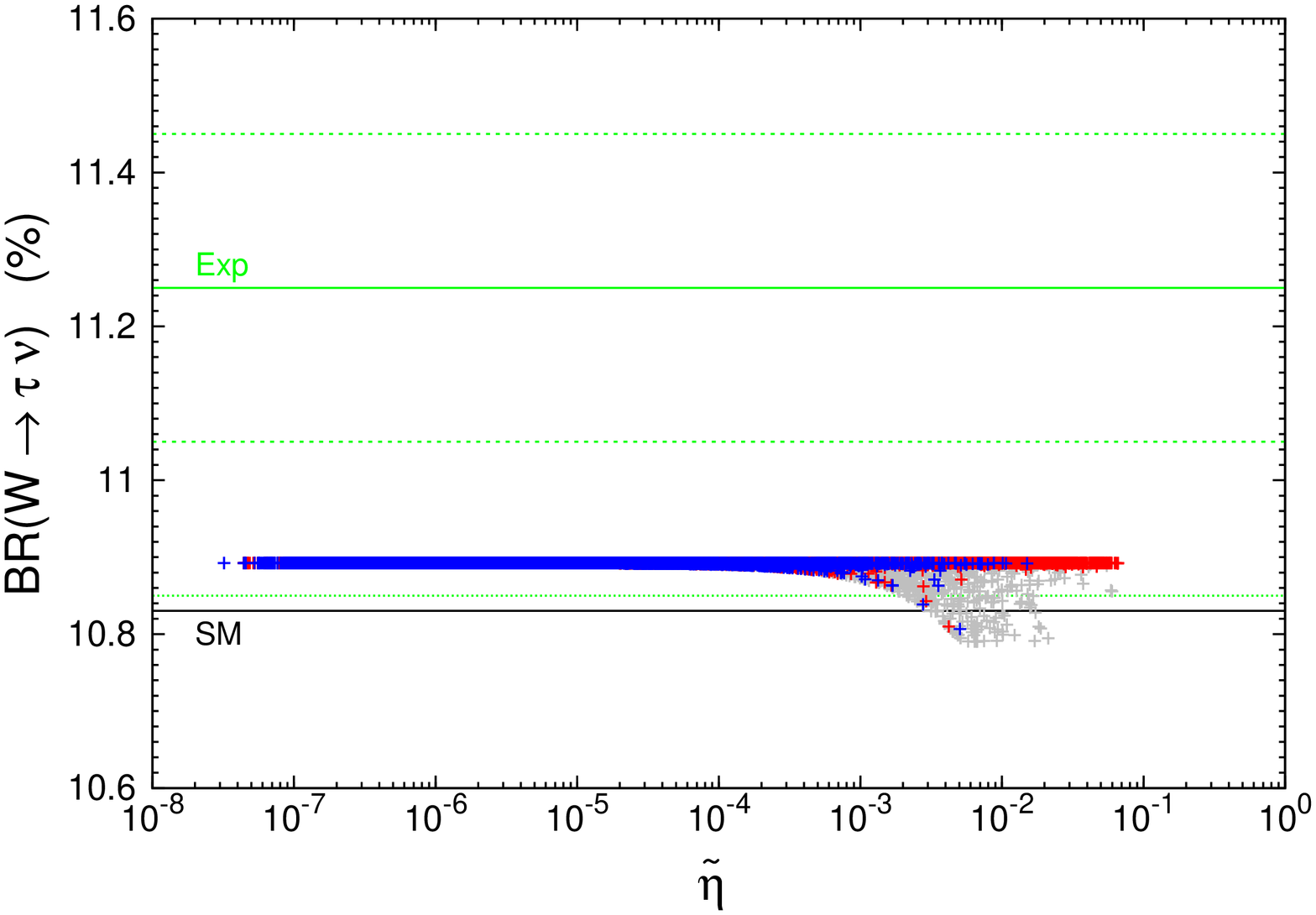,
  width=58mm, clip=, angle=270} 
\end{tabular}
\caption{BR($W \to \ell \nu$) as a function of $\tilde \eta$.
Colour code as in Fig.~\ref{fig:gammaznunu:eta}. Black lines
denote the SM prediction, green ones correspond to the experimental
measurement (full/dashed/dotted corresponding to
central value, $1 \sigma$ and $2 \sigma$ intervals, respectively).
}\label{fig:BRWellnu:etau}
\end{center}
\end{figure}

\medskip
We now proceed to address the impact of sterile neutrinos and the
associated non-unitarity of the $\tilde U_\text{PMNS}$ matrix on its
most directly related observable: leptonic $W$ decays.  As mentioned
in Section~\ref{sec:Wdecays}, there is at present a tension between
the experimental determination of BR($W \to \ell \nu$) and the SM 
expectation, see Eqs.~(\ref{eq:Wellnu:e} - \ref{eq:Wellnu:tau}).  In
view of the sizable deviations that light sterile neutrinos can induce
to (virtual) $W$ mediated processes, we thus explore whether the additional states
 can have an impact on the decay BRs as well.  In
Fig.~\ref{fig:BRWellnu:etau}, we display BR($W \to e \nu$) and BR($W
\to \tau \nu$), both as a function of $\tilde \eta$ (the behaviour of
BR($W \to \mu \nu$) strongly ressembles that of BR($W \to e \nu$), and
we so refrain from displaying the corresponding plot). 
As can be seen from the left panel of Fig.~\ref{fig:BRWellnu:etau},
and similar to what occurred for the $Z$-decay width, non negligible
contributions could indeed soften the tension between the SM
prediction and the experimental values; however, the corresponding
regime is excluded by current MEG bounds. In any case, a
simultaneous reconciliation of the tension 
for the three leptonic $W$ branching ratios  would not have been
possible since, as can be seen from the right panel of
Fig.~\ref{fig:BRWellnu:etau}, the non-unitarity of the  PMNS matrix
worsens the discrepancy for BR($W \to \tau \nu$).  
We notice that in the SM limit, corresponding to $\tilde \eta \sim
0$, there is a minor discrepancy between our values for the BRs and
the SM line~\cite{Kniehl:2000rb}, since in our analysis we do not take
 into account higher-order corrections included in the SM prediction. 

\subsection{$\pmb{\tau}$ decays}
\begin{figure}[h!]
\begin{center}
\psfig{file=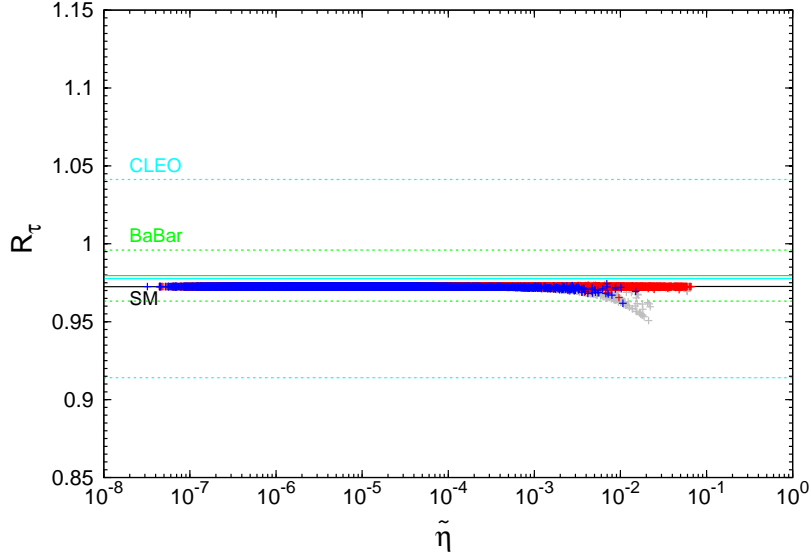, width=80mm,
  clip=, angle=270} 
\caption{$R_\tau$ as a function of $\tilde \eta$.
Colour code as in Fig.~\ref{fig:gammaznunu:eta}. In this case, 
green/cyan lines denote the BaBar/CLEO experimental measurements 
(full and dashed lines corresponding to central values and $1 \sigma$
interval, respectively).} 
\label{fig:Rtau}
\end{center}
\end{figure}

The ratio of leptonic $\tau$ decays $R_\tau$, defined in 
Eq.~(\ref{eq:Rtau}), could  also be sensitive to 
deviations from unitarity.  However, as seen in Fig.~\ref{fig:Rtau},
 the explored parameter space
induces values for $R_\tau$ which are compatible with experimental CLEO and BaBar 
bounds at the (less than) $1 \sigma$ level. 

\begin{figure}[h!]
\begin{center}
\begin{tabular}{cc}
\psfig{file=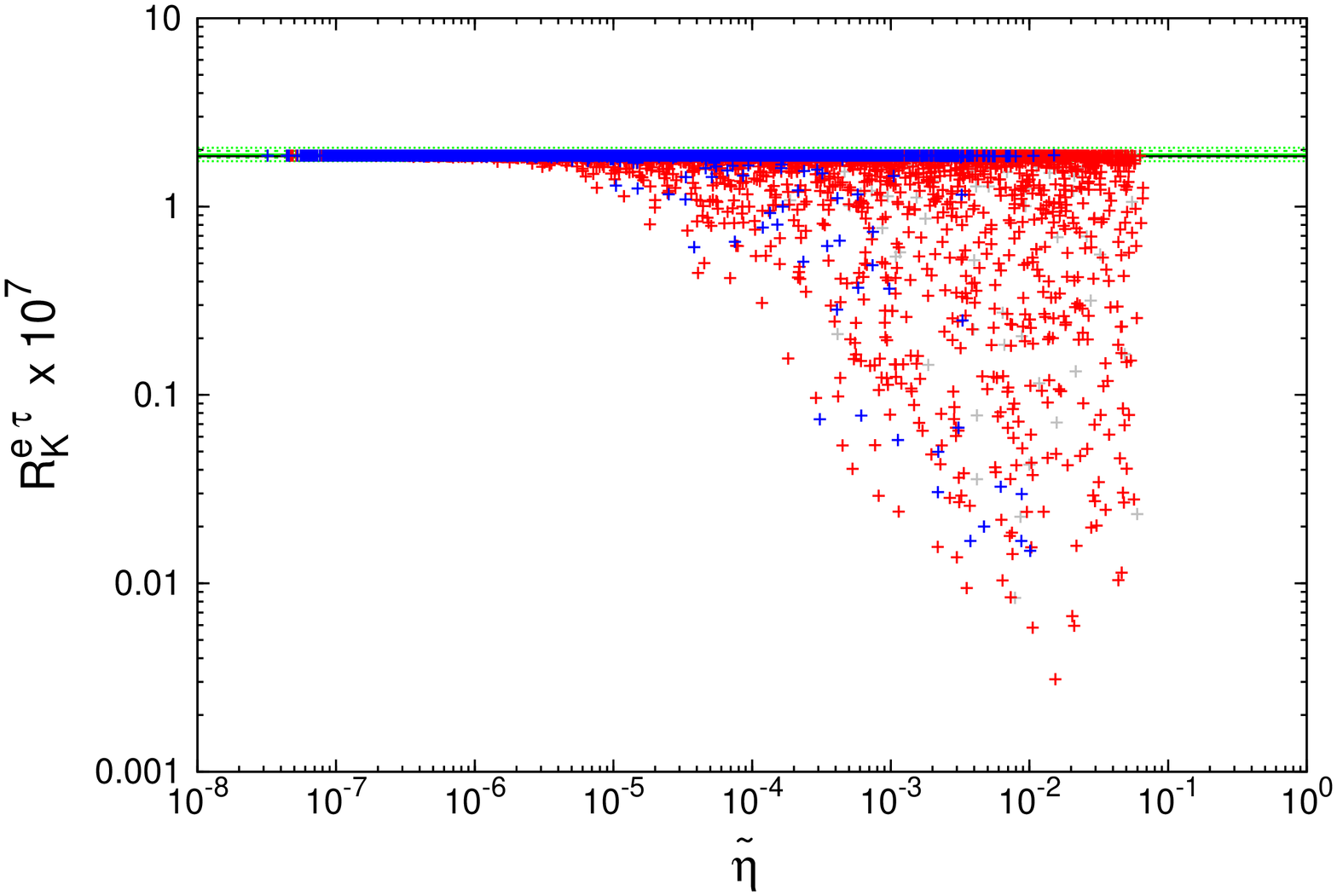, width=55mm, clip=, angle=270} &
\psfig{file=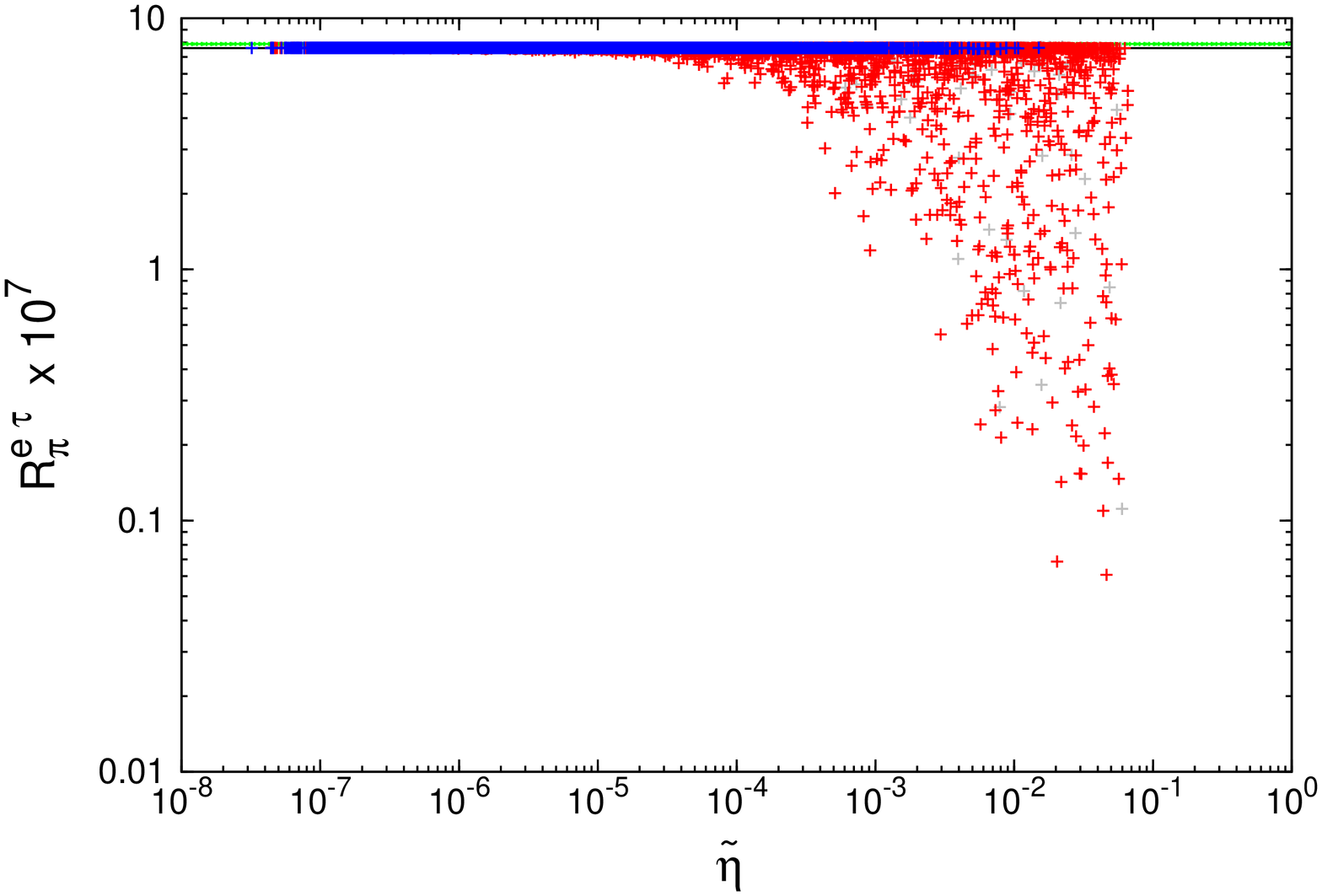, width=55mm, clip=, angle=270}
\end{tabular}
\caption{
Ratios of $\tau$ mesonic and pseudoscalar leptonic decay widths,  
$R_{K, \pi}^{e \tau}$, as a function of $\tilde \eta$.
Line and colour code as in Fig.~\ref{fig:gammaznunu:eta}.
} \label{fig:RKltau}
\end{center}
\end{figure}

\medskip
The ratios $R_K^{e \tau}$ and $R_\pi^{e \tau}$ defined in
Eq.~\eqref{eq:GammaTauKnu:def} might also  be sensitive to deviations from
unitarity. While the SM
predictions are in good agreement with the experimental measurements for
$R_K^{e \tau}$ (which can therefore be used to constrain deviations from
unitarity), there is a discrepancy for $R_\pi^{e \tau}$. 
In Fig.~\ref{fig:RKltau} we display the non-unitarity effects: for 
$R_K^{e \tau}$, and although most of the points lie within the experimental $2\sigma$ interval, 
some exhibit a considerable reduction; a similar situation occurs for
$R_\pi^{e \tau}$ (notice however that the observed deviation  
does not alleviate the aforementioned tension). From both cases, it is
also clear that departures from the SM-like limit mostly occur 
for points already disfavoured by standard cosmology.
We have
also considered $R_K^{\mu \tau}$ and $R_\pi^{\mu \tau}$ and found a
fair agreement between experimental values and theoretical
predictions. 
However, it is important to stress that no single experiment directly
measures $R_K^{\mu \tau}$ or $R_\pi^{\mu \tau}$. 
Moreover, and since some of the measurements
are separated in time by more than five years, there might be systematics
coming from the combination of different experimental results. A detailed
analysis of these possible systematics is beyond the scope of this
work. Nevertheless, a direct experimental measurement of the ratios
$R_P^{\ell \tau}$ would avoid these issues: this could be achieved by
considering the decay chain $\tau \rightarrow P(\rightarrow \ell \nu)
\nu$ and measuring simultaneously $\text{BR}(\tau \rightarrow P
\nu)$ and $\text{BR}(\tau \rightarrow P \nu) \times \text{BR}(P
\rightarrow \ell \nu)$.

\subsection{Leptonic pseudoscalar meson decays}
We now consider the impact of the non-unitarity of the PMNS matrix, as
well as of having light sterile neutrino final states, regarding several
observables related to leptonic pseudoscalar meson decays.

\subsubsection*{Light meson decays -  $\pmb{R_{K,\pi}}$ and 
$\pmb{\Delta r_{K,\pi}}$}
\begin{figure}[h!]
\begin{center}
\begin{tabular}{cc}
\psfig{file=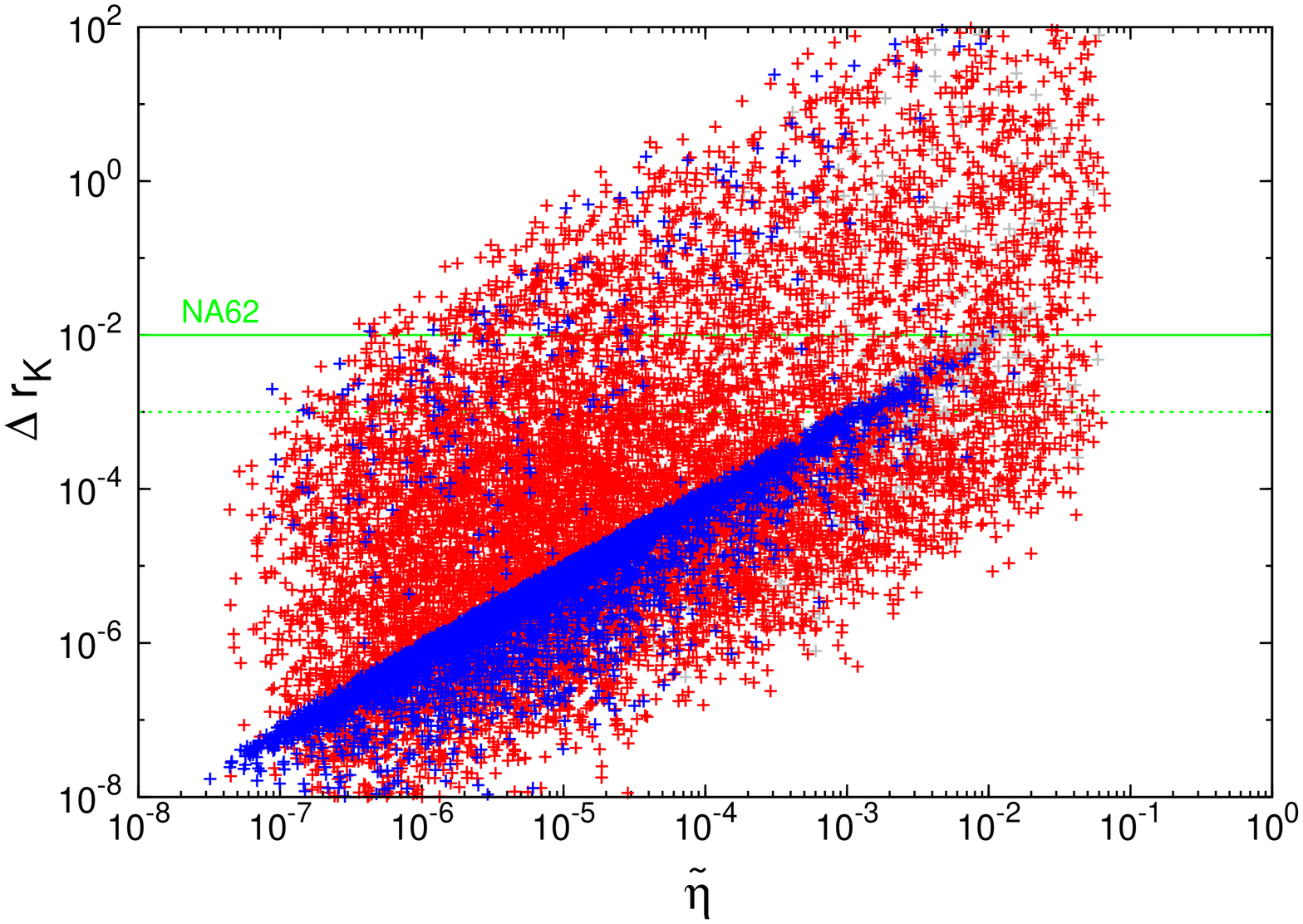, width=55mm, clip=, angle=270} &
\psfig{file=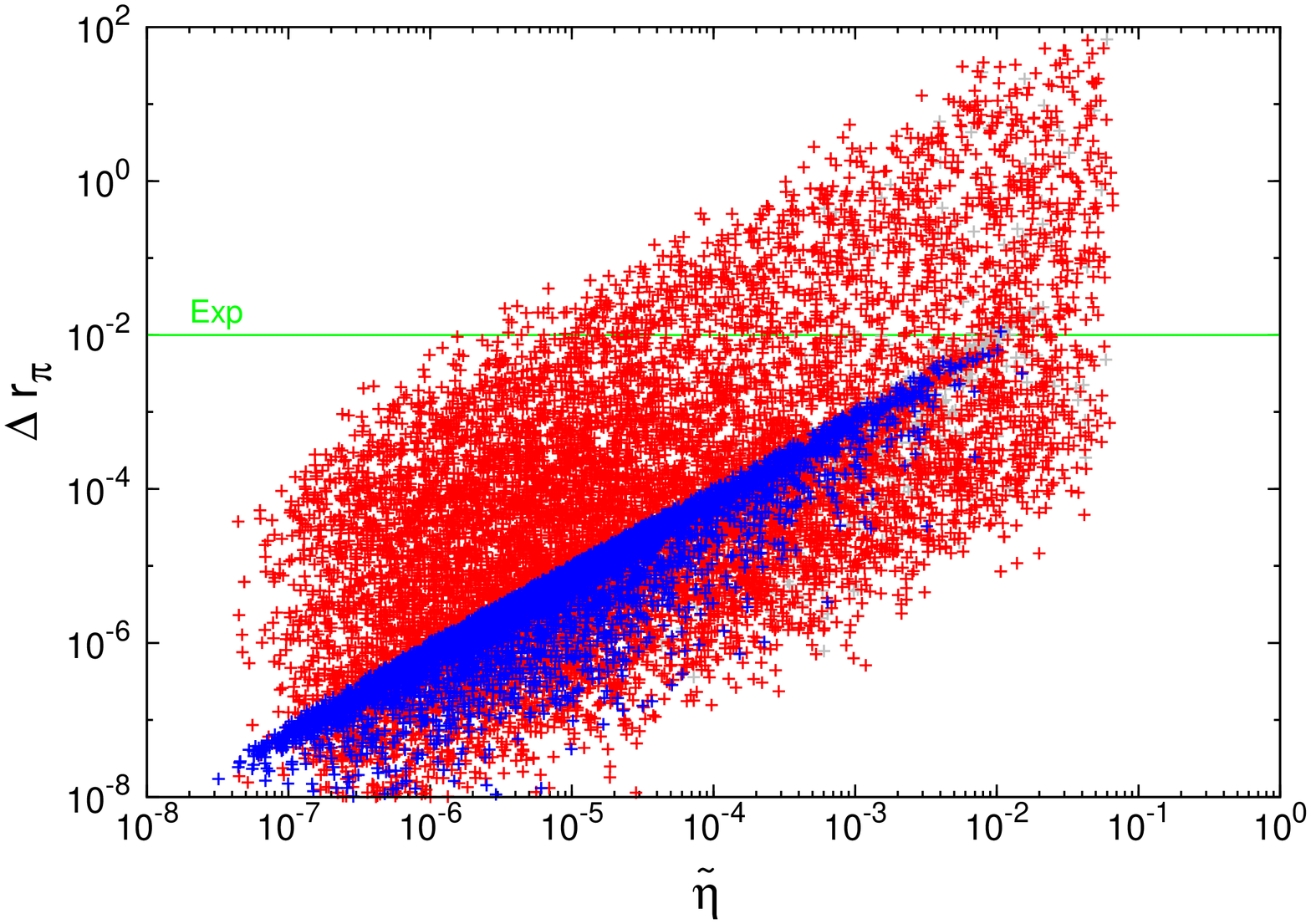, width=55mm, clip=, angle=270}
\end{tabular}
\caption{$\Delta r_K$ and $\Delta r_\pi$ (respectively left and right panels)
as a function of $\tilde \eta$.
Colour code as in Fig.~\ref{fig:gammaznunu:eta}.  
A full green line denotes the experimental upper bound (for $\Delta r_K$
a green dashed line denotes the NA62 expected future sensitivity).}   
\label{fig:DeltaRKPi.eta}
\end{center}
\end{figure}

\begin{figure}[h!]
\begin{center}
\begin{tabular}{cc}
\psfig{file=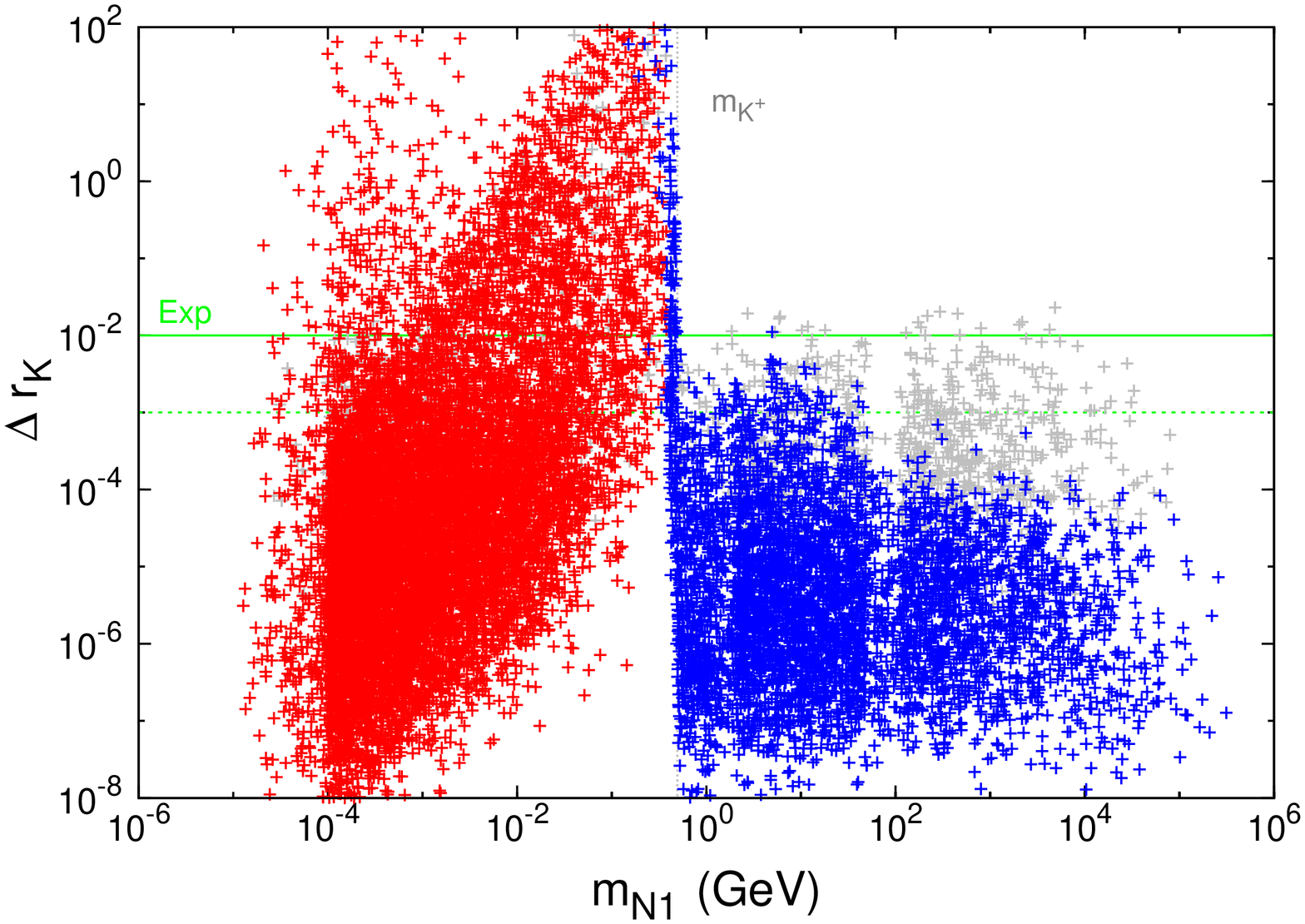, width=55mm, clip=, angle=270} &
\psfig{file=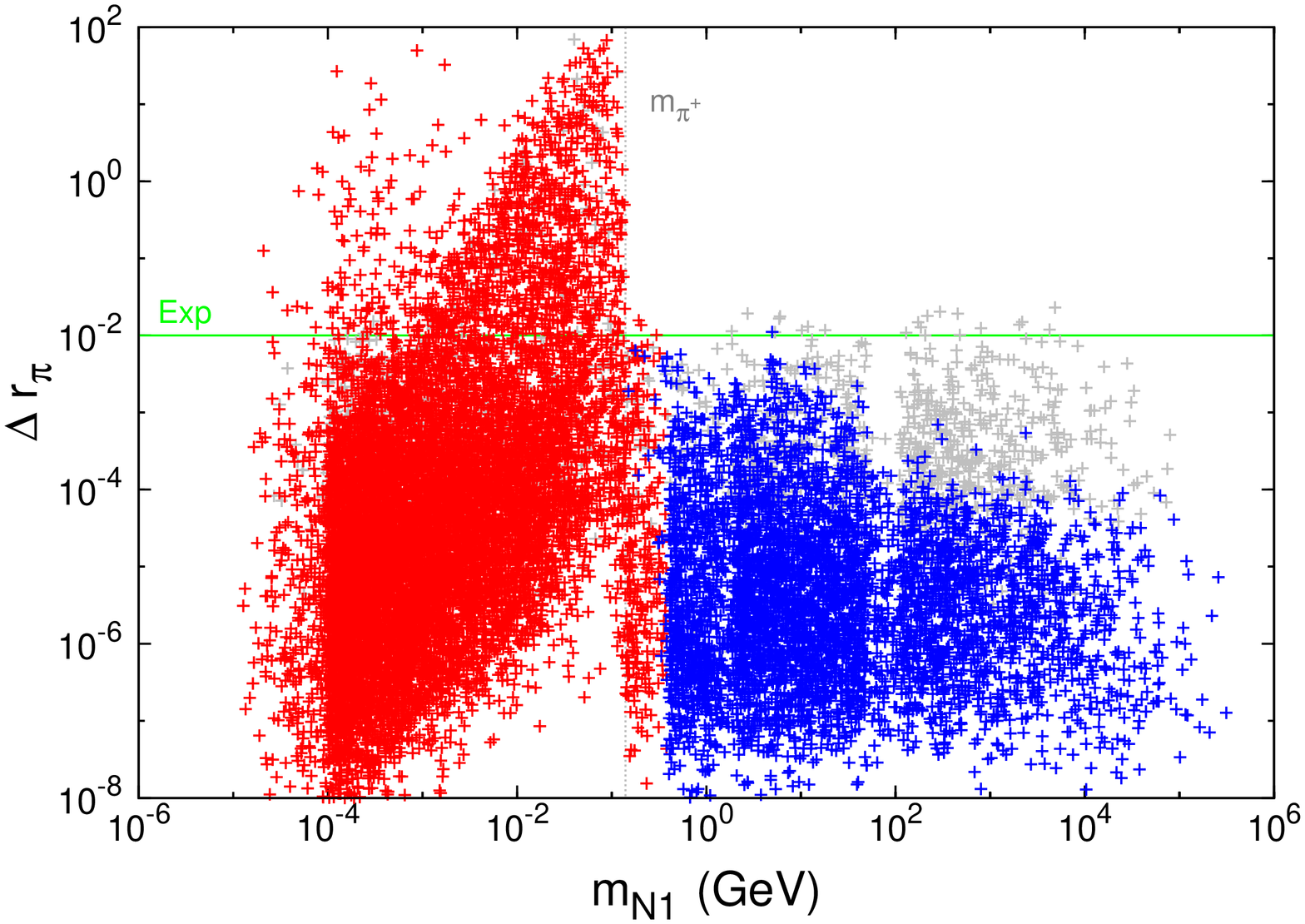, width=55mm, clip=, angle=270}
\end{tabular}
\caption{$\Delta r_K$ and $\Delta r_\pi$ (respectively left and right panels)
as a function of the lightest sterile neutrino mass, $m_{N_1}$.
Colour code as in Fig.~\ref{fig:gammaznunu:eta}.  
Full green lines denote the experimental upper bound (for $\Delta r_K$
a green dashed line denotes the NA62 expected future sensitivity).
Grey vertical dashed lines denote thresholds associated with the
decaying meson mass, respectively $m_{K^+}$ and $m_{\pi^+}$. 
}\label{fig:DeltaRKPi.mN1}
\end{center}
\end{figure}

We begin by discussing the violation of lepton flavour universality in
light pseudoscalar meson decays\footnote{The r\^ole of
  $R_\pi$ in probing non-standard axial and pseudoscalar interactions
  has recently been explored using an effective approach
  in~\cite{Cirigliano:2013xha}.}, parametrised by $\Delta r_{K,\pi}$ as 
defined in Eq.~(\ref{eq:deltar:P}). We display the results in
Figs.~\ref{fig:DeltaRKPi.eta} and~\ref{fig:DeltaRKPi.mN1}. 

Even under the strong constraints arising from the recent MEG bound, one
still recovers the results formerly obtained in~\cite{Abada:2012mc}:
as seen in Fig.~\ref{fig:DeltaRKPi.eta},
large deviations from the SM prediction, within experimental
sensitivity, can be found.  

Figure~\ref{fig:DeltaRKPi.mN1} also offers a clear insight into the
different thresholds related to the decaying meson mass, and the
associated source of deviation from the SM: as discussed in
Section~\ref{sec:lepton.meson.decays}, and as explicitly shown in
Eq.~\eqref{eq:FG}, for sterile neutrinos lighter than the decaying
  meson, one can have sizable deviations from the SM, since phase
  space factors considerably enhance the effects of any deviation
from unitarity of $U_\text{PMNS}$ 
(even if in some case unitarity 
  can be approximately recovered).
For kaons, and contrary to what occurs for the (lighter)
pions, the phase space enhancement 
is such that points in agreement with standard cosmology (blue) can have an
associated $\Delta r_{K} \sim \mathcal{O}(10^{2})$.

\begin{figure}[h!]
\begin{center}
\psfig{file=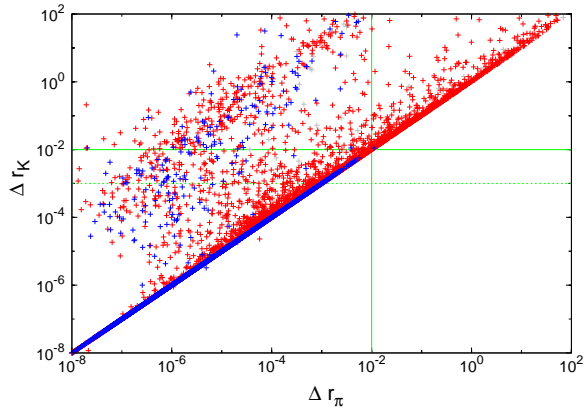, width=58mm,
  clip=, angle=270}  
\caption{$\Delta r_K$ versus $\Delta r_\pi$.
Colour code as in Fig.~\ref{fig:gammaznunu:eta}. 
Full green lines denote the experimental upper bounds (for $\Delta r_K$ a green
dashed line denotes the expected future sensitivity).
} \label{fig:DeltaRK.DeltaRPi}
\end{center}
\end{figure}

Interesting information can also be drawn from analysing the
correlated behaviour of these two observables, $\Delta r_{K}$ and $\Delta r_{\pi}$.
This is displayed in Fig.~\ref{fig:DeltaRK.DeltaRPi}, which exhibits
two interesting characteristics. The first one is
 that many
points are grouped on the diagonal line, thus corresponding to a
scenario where $\Delta r_K$ and $\Delta r_\pi$ are correlated. This is
typically the case when sterile neutrinos are not kinematically
accessible and the deviation is only due to the non-unitarity of
$\tilde U_\text{PMNS}$, since the contribution from non-universality would
be the same for both observables. Secondly, and concerning the 
size of the deviations, notice that  $\Delta r_K$ is 
always larger than $\Delta r_\pi$. This can be understood from the fact
that for certain regimes, phase-space enhancements are possible 
for $R_K$ and not for $R_\pi$. 
Such a result is particular to the ISS scenario. 
In models where the violation of lepton flavour
universality is due to new charged Higgs interactions, one expects much
larger deviations in $K$ decays than in $\pi$ decays (for example, 
in the case of supersymmetric
models~\cite{Masiero:2005wr,RamseyMusolf:2007yb,
Masiero:2008cb,Ellis:2008st,Girrbach:2012km,Fonseca:2012kr}).

\subsubsection*{Light meson decays -  $\pmb{R_{e,\mu}}$ and 
$\pmb{\Delta r_{e,\mu}}$}
\begin{figure}[h!]
\begin{center}
\begin{tabular}{cc}
\hspace*{-7mm}
\psfig{file=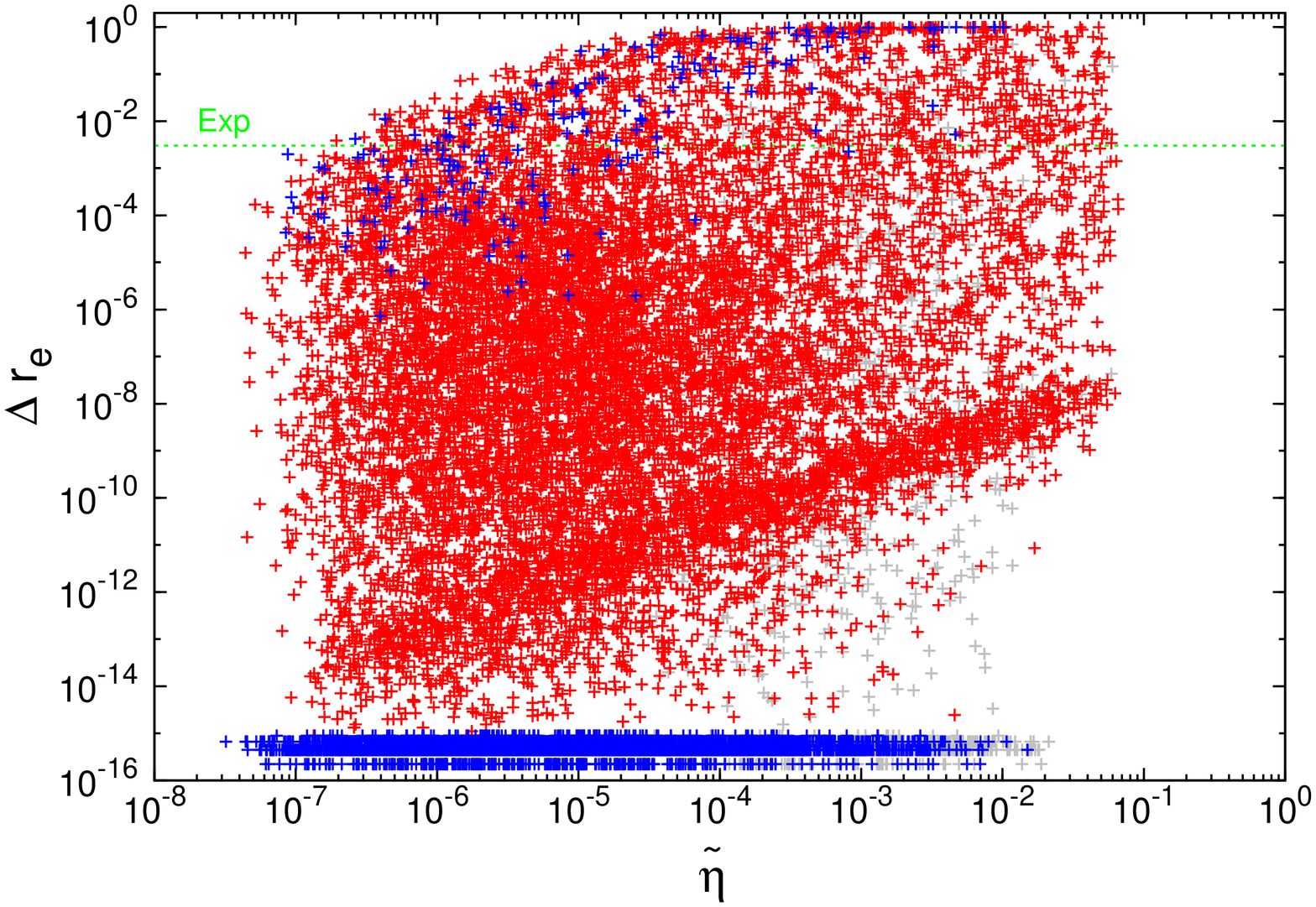, width=58mm, clip=, angle=270} &
\psfig{file=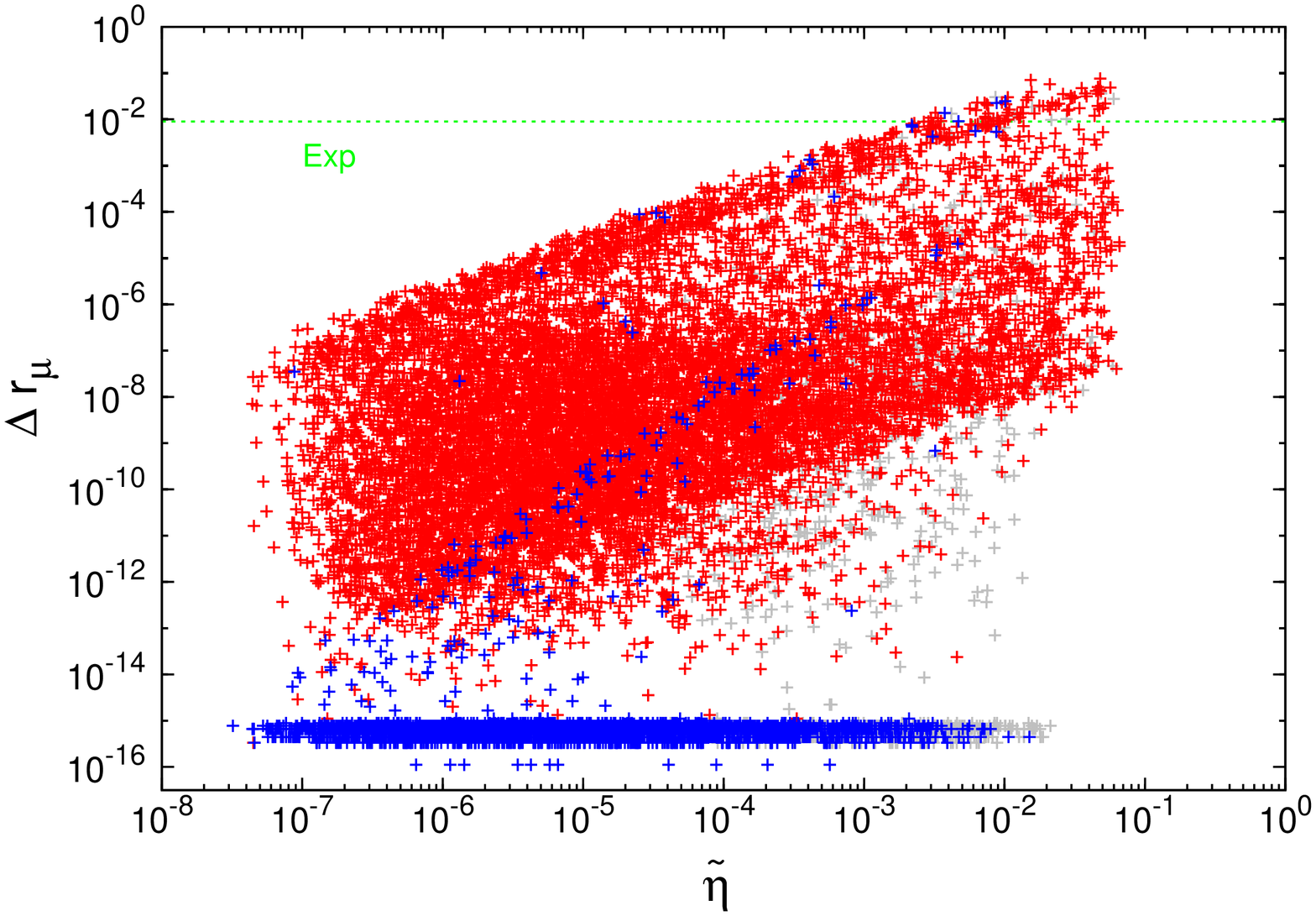, width=58mm, clip=, angle=270}
\end{tabular}
\caption{Ratios of leptonic light meson decay widths, 
$\Delta r_{e}$ and $\Delta r_{\mu}$, as a function of $\tilde \eta$.
Colour code as in Fig.~\ref{fig:gammaznunu:eta}.  
Full green lines denote the experimental upper bounds.
}\label{fig:DeltaRemu.eta} 
\end{center}
\end{figure}

In Fig.~\ref{fig:DeltaRemu.eta}, we present the predictions  regarding the
observables ${\Delta r_{e,\mu}}$ introduced in
Section~\ref{sec:lepton.meson.decays}.  As can be seen from these
figures, in the ISS scenario, especially in the regime of very light
sterile neutrinos, one can indeed easily saturate 
the current experimental upper bound for $\Delta
r_{e}$. However, 
saturating the experimental upper bound on $\Delta
r_{\mu}$ is impossible in the regions of parameter space investigated
in our analysis, except for some very specific points, which are
mostly excluded by cosmological observations. 
For the heavy sterile regime (also corresponding to
the cosmologically viable points), one in general recovers the SM limit.

\subsubsection*{Charmed meson decays - $\pmb{R_{D_s}}$}
\begin{figure}[h!]
 \begin{center}
 \psfig{file=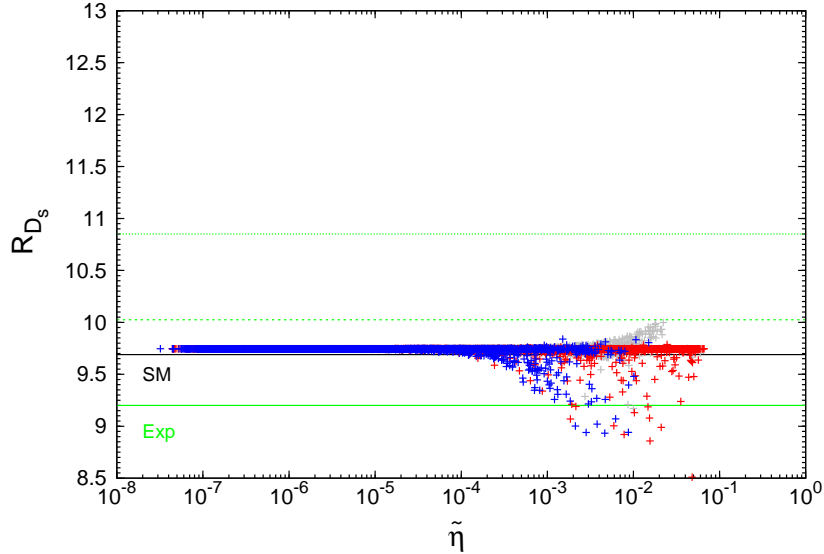, width=80mm, clip=, angle=270} 
 \caption{$R_{D_s}$ as a function of $\tilde \eta$.
Line and colour code as in Fig.~\ref{fig:gammaznunu:eta}.  
 } \label{fig:RDs}
 \end{center}
 \end{figure}

For completeness, we include  in our analysis the
predictions for $R_{D_s}$ 
in the presence of sterile neutrinos\footnote{This observable, as
well as the analogous ratio for $D$ mesons, has been studied in
  Ref.~\cite{Wang:2013hka}. Although we have a similar approach, our
  results differ from the ones obtained in that study, 
  due to some discrepancies in the analytical formulae.}, displaying
the results in Fig.~\ref{fig:RDs}. Current
experimental measurements~\cite{Beringer:1900zz} are compatible with
the SM prediction at the $1 \sigma$ level, as is most of the
parameter space here analysed. Interestingly, in this case the
deviations from unitarity induced by the additional sterile states,
increase the agreement between the ISS theoretical predictions and
experimental observations.

\subsubsection*{Charmed meson decays - $\pmb{R_{D_s}^D}$}
\begin{figure}[h!]
\begin{center}
\begin{tabular}{cc}
\hspace*{-7mm}
\psfig{file=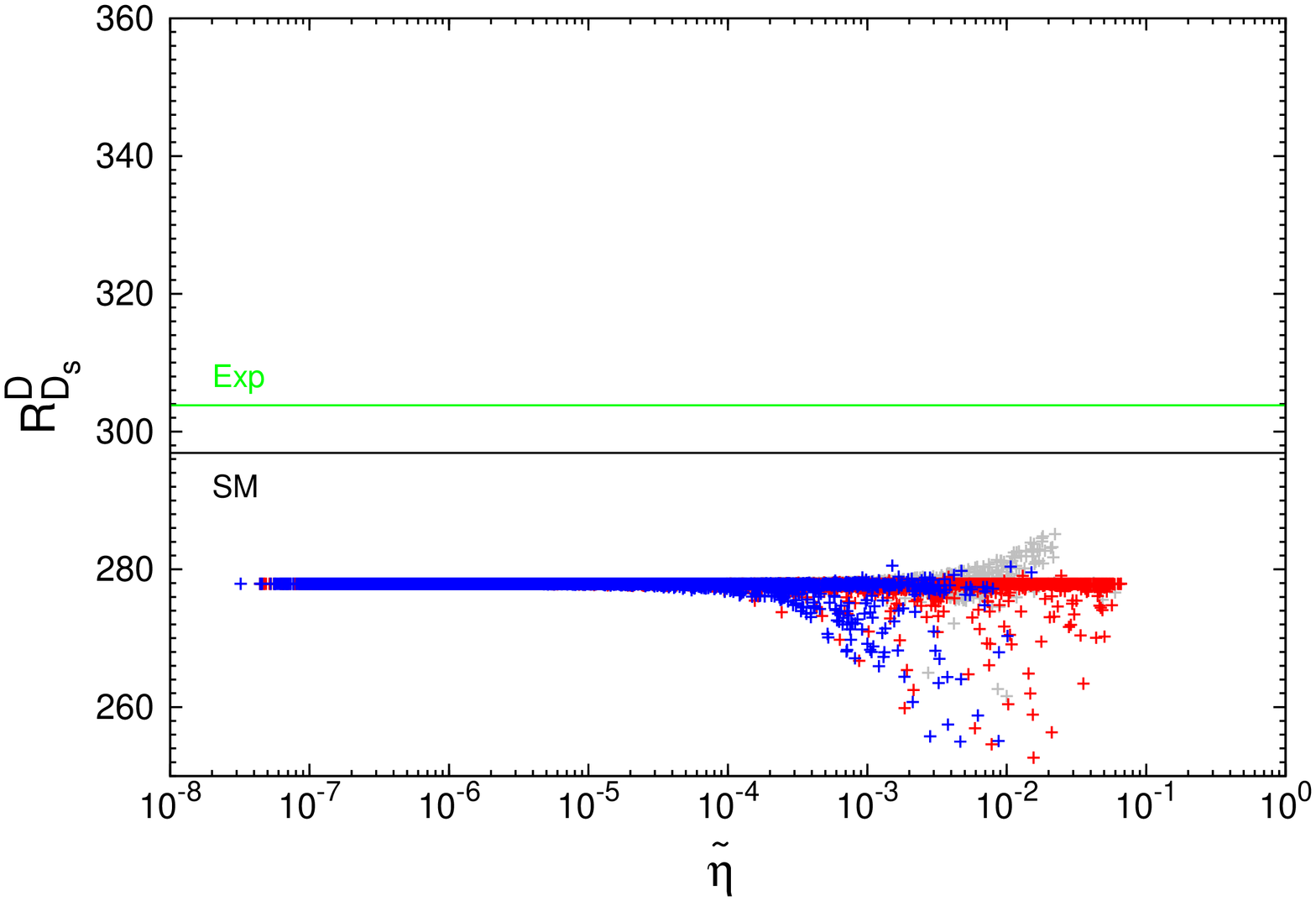, width=58mm, clip=, angle=270} 
 &
\psfig{file=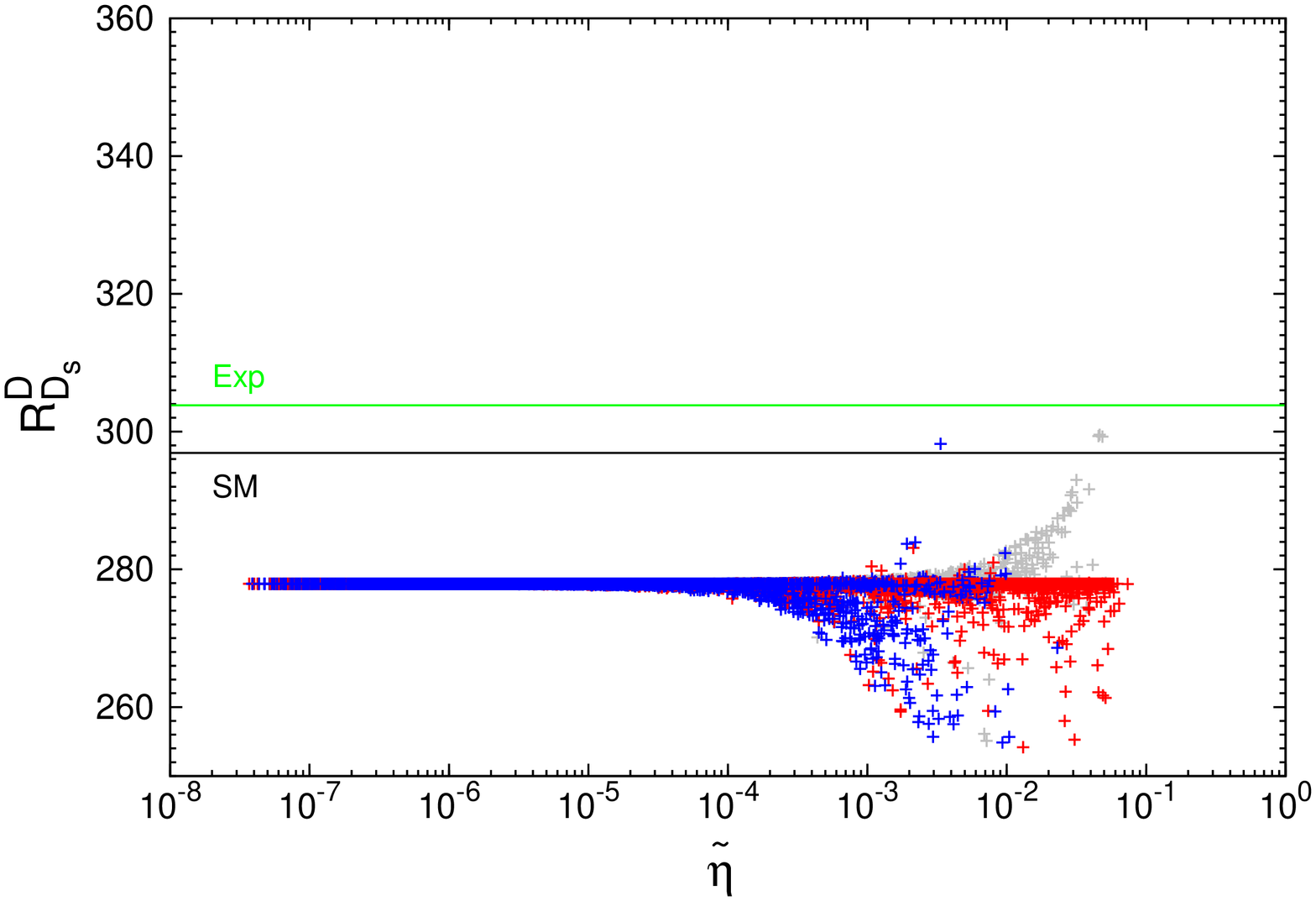, width=58mm, clip=, angle=270}
\end{tabular}
\caption{$R^D_{D_s}$ as a function of $\tilde \eta$. We also display
  the effect of the Dirac phase in the PMNS matrix: 
  $\delta=0$ (left) and randomly varied values, $\delta \in [0,2\pi]$ (right).
Line and colour code as in Fig.~\ref{fig:gammaznunu:eta}.
} \label{fig:RDDs}
\end{center}
\end{figure}

We consider now the impact of the modified $W \ell \nu$ vertex for the
ratio ${R_{D_s}^D} = \Gamma(D_s \to \tau \nu) /\Gamma(D \to \mu
\nu)$. The results of our analysis, displayed as a function of 
$\tilde \eta$, are
collected in Fig.~\ref{fig:RDDs}. In this case we also consider the
effect of a
non-vanishing CP Dirac phase, $\delta$. 
As seen from Fig.~\ref{fig:RDDs}, there is an offset between the SM
expectation and our predictions in the limit $\tilde
\eta \ll 1$ (where one recovers unitarity of the PMNS matrix). This is due to
the fact that the hadronic parameters taken into our computation (see
Eqs.~(\ref{eq:ratiofDsD}, \ref{eq:ratiofKPi})) make use of new
values for the ratio of  $f_{D_s}$ and $f_{D}$, determined from
Lattice QCD~\cite{Becirevic:2013mp}, as well as the very precise
experimental determination of $f_K / f_\pi$~\cite{Antonelli:2009ws}.
In view of this, we only present the central values for the
experimental results. 
It is worth mentioning that taking other determinations of the ratio
${f_{D_s}}/{ f_D}$, as for instance the one reported
in~\cite{Aoki:2013ldr}, would translate into an overall (positive)
correction of around 2\%, which would have little impact on our
phenomenological conclusions.  

While for the previous observables a non-vanishing $\delta$ had a
negligible impact, in the right panel of Fig.~\ref{fig:RDDs} one can
see that for $\delta \neq 0$ a small number of points do succeed in
alleviating the tension between theoretical predictions and
experimental values - provided the above mentioned offset is indeed
accounted for. A larger number of points would have indeed alleviated
the tension, had they not been excluded by the recent MEG bound.

\subsubsection*{$\pmb{B}$ meson decays - BR($\pmb{B\to \tau \nu}$)}
We have also
considered the impact of the modified lepton charged current vertex
regarding leptonic $B$ decays. As mentioned in
Section~\ref{sec:lepton.meson.decays}, this observable suffers from 
uncertainties associated with the determination of hadronic matrix
elements (the $B$ meson decay constant, $f_B$) 
and $V_\text{CKM}^{ub}$, as well as from experimental 
errors~\cite{Adachi:2012mm,Aubert:2007xj}; in the absence of
available data on $B \to (e,\mu) \nu$ decays, one cannot study a ratio
of decay widths to test lepton flavour universality in $B$-meson
decays. 
Our predictions\footnote{Notice that the present computation of 
BR($B\to \tau \nu$) corresponds
to taking the central theoretical values for the different input
parameters; due to the size of the theoretical error band, there is a
significant overlap between the experimental and the theoretical $1
\sigma$ intervals. } (based on the input values
for $f_B$ and $V_\text{CKM}^{ub}$ given in
Section~\ref{sec:lepton.meson.decays}) for BR($B\to \tau \nu$) in
the framework of the ISS, correspond to the SM theoretical prediction 
(within the \% level). Thus, we do not display any plots for this observable.

\subsection{$\pmb{B^\pm \to D \ell \nu}$ meson decays}
\begin{figure}[h!]
\begin{center}
\begin{tabular}{cc}
\hspace*{-7mm} 
\psfig{file=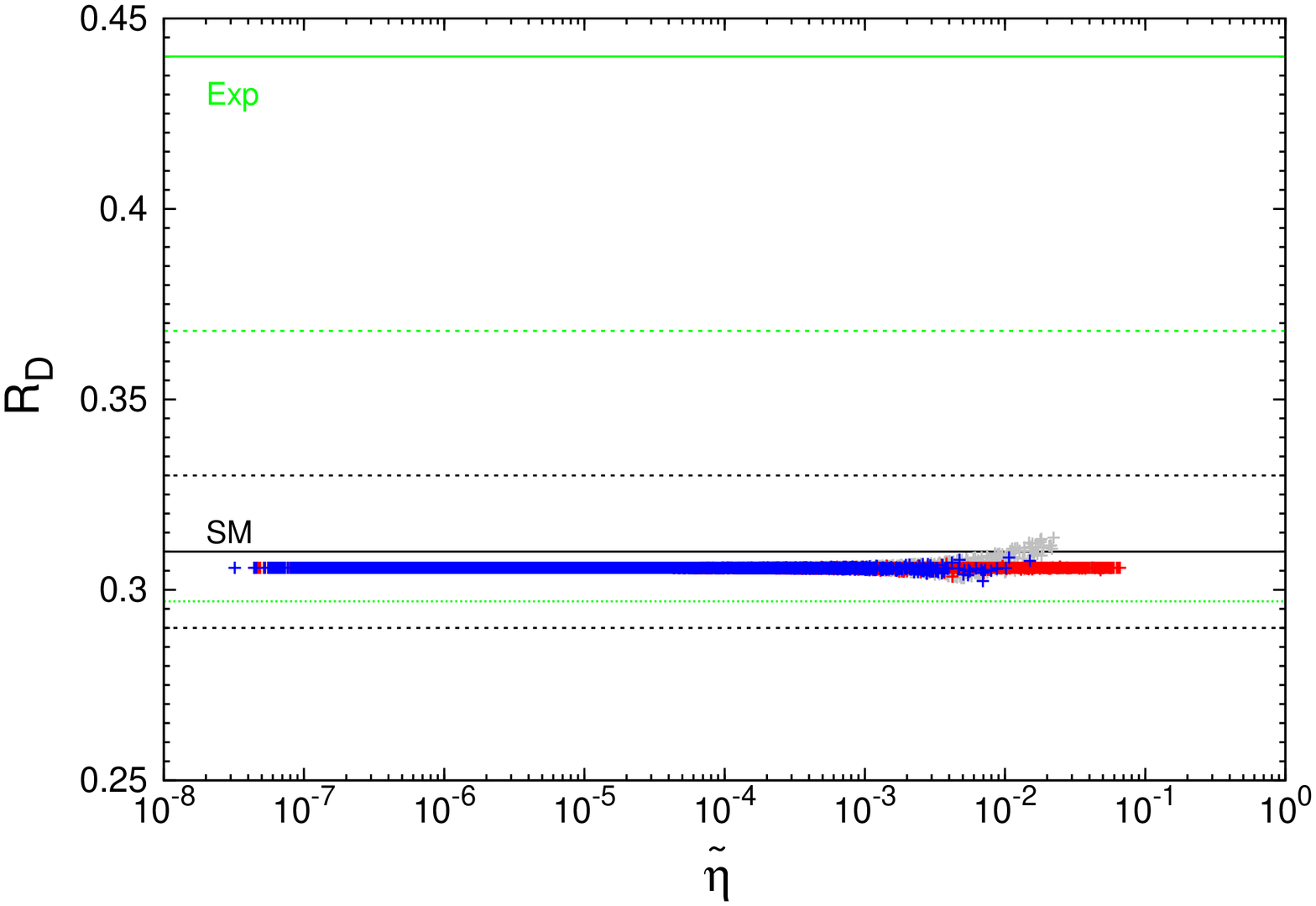, width=58mm, clip=, angle=270} &
\psfig{file=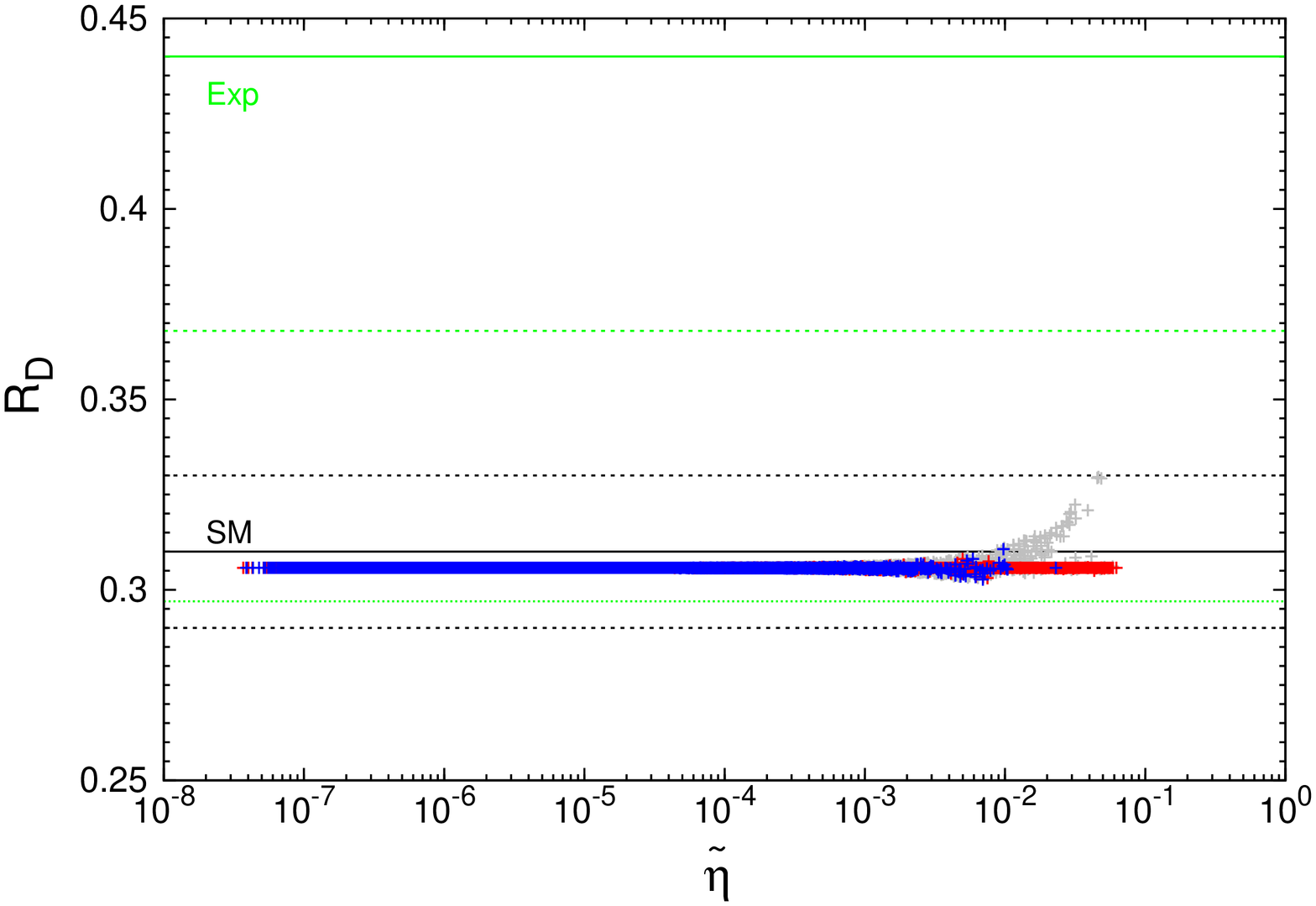, width=58mm, clip=, angle=270}
\end{tabular}
\caption{$R_{D}$ as a function of $\tilde \eta$, displaying 
the effect of the Dirac phase in the PMNS matrix: 
$\delta=0$ (left) and randomly varied values $\delta \in [0,2\pi]$ (right).
Line and colour code as in Fig.~\ref{fig:gammaznunu:eta}.
}\label{fig:BtoDellnu} 
\end{center}
\end{figure}

We finally address the semileptonic decays of the charged $B$ meson 
into a neutral $D$ and a lepton pair, in particular the ratio $R(D)$,
defined in Eq.~(\ref{eq:RDSL:def}). In view of the non-negligible
contributions of the modified $W \ell \nu$ vertex (due to the
presence of sterile neutrinos), we now investigate if the present framework could
alleviate the existing tension between the SM prediction and the
recent bounds (cf. Eqs.~(\ref{eq:RDSL:exp}, \ref{eq:RDSL:th})).
In Fig.~\ref{fig:BtoDellnu}, we display $R(D)$ as a function of the
non-unitarity parameter $\tilde \eta$ and, as done for $R^D_{D_s}$,
we also illustrate the
effect of a non-vanishing CP Dirac phase,~$\delta$.

As one can see from Fig.~\ref{fig:BtoDellnu}, although the ISS could
potentially give rise to contributions to $R(D)$ providing a minor
alleviation   
of the existing tension between SM predictions and
experimental measurements (especially in the 
case of non-vanishing CP Dirac phases), these are excluded due to the strong
constraints arising from recent MEG bounds.

Other experimental measurements are expected in the future, and these
will perhaps soften the deviation from the theoretical 
estimations\footnote{A huge effort is currently being made regarding 
the determination of $B$ meson semileptonic decay form factors, 
as can be noticed from the 
  FLAG review~\cite{Aoki:2013ldr}. Very recent studies, after
  completion of our numerical analysis, have been
  reported~\cite{Atoui:2013ksa}, but the results do not change the
  conclusions of this work.}.   
Moreover, observables related to semileptonic 
$B$ decays are not free from QCD uncertainties (form factors), 
while such was not the case for other observables here studied.
Finally, sources of NP in the lepton sector, other than  
the minimal inverse seesaw scenario
studied here, can be considered.  

\section{Conclusions}\label{sec:concs}
In this work we have tried to reconcile theory and experiment in
leptonic and semileptonic decays, under the hypothesis of 
New Physics contributions associated with the lepton sector.   
We have considered tree-level corrections to the SM charged current
interaction $W \ell \nu$ vertex, due to the presence of sterile
neutrinos (right-handed, or singlet components) which arise in several
extensions of the SM aiming at addressing neutrino mass generation.

The phenomenological implications of these extensions are vast, and there are 
presently strong experimental and observational bounds (from
laboratory, cosmology, as well as from electroweak precision tests) on
the mass regimes and on the size of the active-sterile mixings. 

In our analysis we have focused on the impact of 
the additional states for leptonic charged currents: the modification
of the Standard Model  $W \ell \nu$ vertex can
lead to potentially large contributions to observables involving one
or two neutrinos in the final state. We have derived 
complete analytical expressions for all the observables in the
framework of the SM extended by sterile states, taking into account
massive leptons and their mixings.
In order to illustrate the impact of the sterile fermions, we have
considered the framework of the inverse seesaw mechanism.

Although conducted for a specific seesaw realisation,
our analysis reveals that New Physics in the lepton sector - in the
form of additional sterile states - can indeed lead to 
contributions to some of the leptonic and
semileptonic decays here considered ($\tau$ leptonic and mesonic
decays, leptonic $ \pi,\ K,\ D, \ D_s$ decays and semileptonic $B\to
D\ell \nu$ decays). Notice however that these are accompanied by
sizable contributions to rare radiative lepton decays: in particular, 
new MEG bounds on BR($\mu \to e \gamma$) preclude important ISS
contributions which would otherwise allow to alleviate the tension
between theory and experiment.
We extended our analysis to observables which
are likely to be studied in the near future (for example $R^{\ell
  \tau}_K $), predicting their expected range for the
investigated parameter space. 

Our analysis reveals that, of the different investigated observables,  
$R_{K,\pi}$ are clearly the most powerful ones in constraining the model.
Contrary to other observables, $R_P$ is helicity suppressed in the SM,
and as a consequence very small values are predicted. 
Any SM extension where helicity suppression is no longer
present (or is at least alleviated) should then allow
for sizable deviations in $R_P$, as is the case of the ISS scenario
 addressed in this work. 

\section*{Acknowledgements}
We are grateful to Damir Becirevic for many useful and enlightening
discussions. A. V. thanks Tommaso Spadaro for his insight on the
experimental perspectives of the NA62 experiment.  This work has been
partly done under the ANR project CPV-LFV-LHC NT09-508531. 
The work of C.~W. is suppported by the Spanish MINECO under grant FPA-2012-31880. The
authors acknowledge partial support from the European Union FP7 ITN
INVISIBLES (Marie Curie Actions, PITN-\-GA-\-2011-\-289442).

\end{document}